\documentclass[12pt,preprint]{aastex}
\begin{document}

\title{Mg II Absorption Systems in SDSS QSO Spectra\altaffilmark{1}}

\author{Daniel B. Nestor\altaffilmark{2}, David
A. Turnshek\altaffilmark{2}, and Sandhya M. Rao\altaffilmark{2}}

\affil{Department of Physics \& Astronomy, University of Pittsburgh,
Pittsburgh, PA 15260}

\altaffiltext{1}{Based on data obtained in the Sloan Digital Sky
Survey.}

\altaffiltext{2}{email: dbn@astro.ufl.edu;
turnshek@quasar.phyast.pitt.edu; rao@everest.phyast.pitt.edu}

\begin{abstract}
We present the results of a \ion{Mg}{2} $\lambda\lambda2796,2803$
absorption-line survey using QSO spectra from the Sloan Digital Sky
Survey Early Data Release.  Over 1,300 doublets with rest equivalent
widths greater than 0.3\,\AA\ and redshifts $0.366 \le z \le 2.269$
were identified and measured.   We find that the $\lambda2796$ rest
equivalent width ($W_0^{\lambda2796}$) distribution is described  very
well by an exponential function $\partial N/\partial W_0^{\lambda2796}
= \frac{N^*}{W^*}\,e^{-\frac{W_0}{W^*}}$, with $N^*=1.187 \pm 0.052$
and $W^*=0.702 \pm 0.017$\AA.  Previously reported power law fits
drastically over-predict the number of strong lines.  Extrapolating
our exponential fit under-predicts the number of $W_0^{\lambda2796}
\le 0.3$\,\AA\ systems,  indicating a transition in $\partial
N/\partial W_0^{\lambda2796}$ near $W_0^{\lambda2796} \simeq
0.3$\,\AA.  A combination of two exponentials reproduces the observed
distribution well, suggesting that \ion{Mg}{2} absorbers are the
superposition of at least two physically distinct populations of
absorbing clouds.   We also derive a new redshift parameterization for
the number density of $W_0^{\lambda2796} \ge 0.3$\,\AA\ lines:
$N^*=1.001\pm0.132\,(1+z)^{0.226\pm0.170}$ and
$W^*=0.443\pm0.032\,(1+z)^{0.634\pm0.097}$\,\AA.  We find that the
distribution steepens with decreasing redshift, with $W^*$ decreasing
from $0.80\pm0.04$\,\AA\ at $z = 1.6$ to $0.59\pm0.02$ \,\AA\ at
$z=0.7$.   The incidence of moderately strong ($0.4\,\mathrm{\AA}
\lesssim W_0^{\lambda2796} \lesssim 2$\,\AA) \ion{Mg}{2} $\lambda2796$
lines does not show evidence for evolution with redshift.   However,
lines stronger than $\approx 2$\,\AA\ show a decrease relative to the
no-evolution prediction with decreasing redshift for $z \lesssim 1$.
The evolution is stronger for increasingly stronger lines.   Since
$W_0$ in saturated absorption lines is an indicator of the velocity
spread of the absorbing clouds, we interpret this as an evolution in
the kinematic properties of galaxies from moderate to low redshift.
Monte Carlo simulations do not reveal any strong systematic  effects
or biases in our results.  While more recent SDSS QSO spectra  offer
the opportunity to increase the \ion{Mg}{2} absorber sample by another
order of magnitude, systematic errors in line identification and
measurement will begin to dominate in the determination of absorber
property statistics.
\end{abstract}


\section{Introduction}
Intervening QSO absorption lines provide an opportunity to study the
evolution of  galaxies selected via their gas cross-sections,
independent of their stellar luminosities, from the earliest era of
galaxy formation up to the present epoch.   The chance alignment of
galactic gas with the line of sight to a distant QSO allows the unique
opportunity to study the gaseous phase of galaxies, without the biases
involved in luminosity limited studies.  In particular, low-ion metal
lines and  strong \ion{H}{1} lines can be used to sample the
low-ionization and neutral gas  bound in galactic systems.  Surveys
for intervening \ion{Mg}{2} absorption systems are particularly
useful, since the \ion{Mg}{2} doublet at rest wavelengths
$\lambda\lambda2796,2803$ allows ground based  surveys to track them
down to relatively low redshift $(z=0.11)$.  Strong \ion{Mg}{2}
systems are good tracers of large columns of neutral hydrogen gas (Rao
\& Turnshek 2000)  and are useful probes of the velocity structure of
the neutral gas components of galaxies, including damped Ly$\alpha$
(DLA) systems.  Furthermore, unsaturated low-ionization metal lines
associated with strong \ion{Mg}{2} systems can be used to deduce
neutral phase metal abundances (Pettini et al. 1999;  Nestor et
al. 2003; Prochaska et al. 2003).

Many surveys have studied the statistical properties of \ion{Mg}{2}
absorption systems.  Lanzetta, Turnshek, \& Wolfe (1987) presented the
first significant survey and provided the benchmark results for the
statistics of \ion{Mg}{2} absorbers at relatively high redshift ($1.25
< z < 2.15$).  They found marginal evidence for evolution in the
number density of absorbers, but no significant evidence for evolution
of the $\lambda2796$ rest equivalent width ($W_0^{\lambda2796}$)
distribution, which they fit equally well with an exponential and a
power law.  Tytler et al. (1987) and Sargent, Steidel, \&  Boksenberg
(1988) provided data on \ion{Mg}{2} systems at lower redshift $(0.2
\lesssim z \lesssim 1.5)$ and found that the comoving number density
of absorbers\footnote{We use partial derivatives in place of the
traditional $dN/dz$ since, as we discuss in \S3, the $z$ and
$W_0^{\lambda2796}$ dependencies of $N$ are not separable.} ($\partial
N/\partial z$)  increases with redshift, consistent within the errors
with no evolution.  Caulet (1989) compared the Lanzetta, Turnshek, \&
Wolfe results to results at lower redshift, finding more moderately
strong systems and fewer weak systems at $\left<z\right>=1.6$ as
compared to $\left<z\right>=0.5$.  The study by Steidel \& Sargent
(1992, hereafter SS92), which contained 107 doublets with rest
equivalent widths $W_0^{\lambda2796} \ge 0.3$\,\AA\ over the redshift
range $0.23 \le z \le 2.06$, has served as the standard for the
statistics of the distribution of \ion{Mg}{2} $\lambda2796$ rest
equivalent  widths $\partial N/\partial W_0^{\lambda2796}$,  redshift
number density $\partial N/\partial z$, and redshift evolution for the
past decade.  Their conclusions included the following:
\begin{enumerate}
\item{ The $W_0^{\lambda2796}$ distribution can be fit equally well
with either an  exponential or power law.}
\item{$\partial N/\partial z$ for systems with $W_0^{\lambda2796} \ge
  0.3$\,\AA\ increases with redshift in a manner consistent with  no
  evolution.}
\item{The redshift number density $\partial N/\partial z$ for the
strongest lines does show  evidence for evolution, with $\partial
N/\partial z$ decreasing  from the no evolution prediction with
decreasing redshift.}
\item{The mean $W_0^{\lambda2796}$ increases with redshift.}
\item{There is no evidence for correlation between the doublet ratio
$W_0^{\lambda2796} / W_0^{\lambda2803}$ and redshift.}
\end{enumerate}
Churchill et al. (1999, hereafter CRCV99),  presented a study of 30
``weak'' ($W_0^{\lambda2796} \le 0.3$\,\AA) systems,  and determined
$\partial N/\partial W_0^{\lambda2796}$ and $\partial N/\partial z$
for the weak extreme of the \ion{Mg}{2} $W_0^{\lambda2796}$
population.  They favored a power law over an exponential for the
parameterization of $\partial N/\partial W_0^{\lambda2796}$.

The Sloan Digital Sky Survey (SDSS; York et al. 2000) provides an
opportunity to improve the statistics of \ion{Mg}{2} and other low-ion
absorption systems by increasing the number of measured systems by
orders of magnitude.  The statistics and evolution of the absorber
number density, which is the product of the space density and the
absorption cross-section, and the distribution of line strengths,
which correlate with the number of kinematic subsystems and absorber
velocity dispersion,  are important factors in understanding the
physical nature of \ion{Mg}{2} systems and their evolution.
Furthermore, the strong connection to DLA and sub-DLA systems (Rao \&
Turnshek 2000; Rao, Turnshek, \& Nestor 2004, in preparation) makes
the parameterization of the \ion{Mg}{2} properties key to the study of
large \ion{H}{1} column density systems.  Finally, only large surveys
provide statistically significant numbers of rare, ultra-strong
systems for complementary imaging and high-resolution kinematic
studies.

In this paper, we present the results from a \ion{Mg}{2}
$\lambda\lambda2796,2803$ survey using QSO spectra from the SDSS Early
Data Release (EDR; Stoughton et al. 2002).  The results of  our
analysis, using only a small fraction of the final SDSS database, is
of interest because with the large number of \ion{Mg}{2} doublets
identified (over 1,300 with $W_0^{\lambda2796} \ge 0.3$\,\AA),
systematic errors already compete with Poissonian errors in the
measured statistics.  In \S2 we describe the SDSS EDR data set, our
doublet finding algorithm, and tests for systematic biases.  In \S3 we
present the results of our analyses, describe the absorber statistics
and  their parameterizations, and consider sources of systematic
errors.  In \S4 we discuss our interpretations.  We present our
conclusions in \S5.

\section{Analysis}

\subsection{The Data}
The SDSS EDR provides 3814 QSO spectra, approximately 3700 of which
are of QSOs with sufficiently high redshift  ($0.37 \le z_{QSO} \le
5.30$, $\left<z_{QSO}\right> \simeq 1.53$)  to allow detection of
intervening \ion{Mg}{2} $\lambda\lambda2796,2803$ doublets.   We
analyzed all available spectra, regardless of QSO magnitude, as only a
small number were too faint to detect the strongest systems in our
catalog.  This data set combines a large number of sightlines with a
wide spectral coverage: from 3800\,\AA\  at a resolution of 1800, to
9200\,\AA\ at a resolution of 2100.  This corresponds to a \ion{Mg}{2}
absorption redshift coverage of $0.37 \le z_{abs} \le 2.27$.  Typical
signal to noise  ratios in the SDSS EDR QSO spectra are such that
the number of detectable systems drops significantly for
$W_0^{\lambda2796}$ less than $\approx$ 0.6\,\AA,
though our catalog includes systems down to $W_0^{\lambda2796} =
0.3$\,\AA.  The strongest system that we found has $W_0^{\lambda2796}
= 5.68$\,\AA.  At all of these strengths, the \ion{Mg}{2}
$\lambda2796$ line is typically saturated.  Thus, no column density
information can be gleaned from $W_0^{\lambda2796}$.  However, large
$W_0^{\lambda2796}$ does track high \ion{H}{1} columns and exhibits
correlation with line of sight velocity dispersion and metallicity
(see Nestor et al. 2003).   Although the EDR QSO catalog selection
properties are not homogeneous, and therefore  not appropriate for
statistical  analyses, the QSO selection criterion should have little
affect on the analysis of intervening absorption.  Effects related to
the QSO selection that could impact our results are discussed in
\S\ref{sec_se}.

\subsection{Continuum Fitting}
In order to search the spectra for \ion{Mg}{2} doublets, continua were
first fitted to the data using the following algorithm.   Data below
the QSO rest-frame Ly$\alpha$ emission  were excluded to avoid the
unreliable process of searching for lines in the Ly$\alpha$ forest, as
were  data longward of the \ion{Mg}{2} $\lambda\lambda2796,2803$
emission feature.  As our only interest is in detecting and measuring
the strength of absorption features, we defined a continuum to be the
spectrum as it would appear in the absence of absorption features,
i.e., the true continuum plus broad emission lines.  An underlying
continuum was fitted with a cubic spline, and broad emission and broad
and narrow absorption  features were fitted with one or more Gaussians
and subtracted.  This process was iterated several times to improve
the fits of both the spline and Gaussians.  Except for a few
problematic areas  in a small number of spectra, this technique
provided highly satisfactory results.  Note that by avoiding the
Ly$\alpha$ forest, we do not experience the difficulties that arise in
trying to define a continuum in that region.  The fluxes and errors in
flux were then normalized by the fitted  continua, and these
normalized spectra were used for subsequent analyses.

\subsection{\ion{Mg}{2} Doublet Finding Algorithm}
We searched the continuum-normalized SDSS EDR QSO spectra for
\ion{Mg}{2}  $\lambda\lambda2796,2803$ doublet candidates.  All
candidates were interactively checked for false detections,
satisfactory continuum fits, blends with absorption lines from other
systems,  and other special cases.  We used an optimal extraction
method (see Schneider et al. 1993) to measure each $W_0$:
\begin{equation}
(1+z)\,W_0 = \frac{\sum_i P(\lambda_i-\lambda_0)\, (1-f_i)}{\sum_i
P^2(\lambda_i-\lambda_0)}\,\Delta\lambda,
\end{equation} 
\begin{equation}
(1+z)\,\sigma_{W_0} = \frac{\sqrt{\sum_i P^2(\lambda_i-\lambda_0)\,
\sigma_{f_i}^2}}{\sum_i P^2(\lambda_i-\lambda_0)}\,\Delta\lambda,
\end{equation} 
where $P(\lambda_i-\lambda_0)$ represents the line profile, and $f_i$
and $\sigma_{f_i}$ the normalized flux and flux error per pixel.   The
sum is performed over an integer number of pixels that cover at least
$\pm 3$ characteristic Gaussian widths.  Many of the lines found were
unresolved.  For these, it was appropriate to use the line spread
function for the optimal extraction profile, and a Gaussian with a
width corresponding to the resolution generally provided a very
satisfactory fit.  A large proportion of the lines, however, were at
least mildly resolved.  For  these systems, a profile obtained by
fitting a Gaussian whose width is  only constrained to be greater than
or equal to the unresolved width was appropriate.  For the large
majority of the mildly resolved  lines, single Gaussians were
satisfactory fits and our method for measuring $W_0$ gave results
consistent with, and with higher  significance than, other methods
such as direct integration over the line.  In rare cases
($\approx4\%$), a single Gaussian was a poor description of the line
profile, and more complex profiles, such as a displaced
double-Gaussian, were employed.

Identifying \ion{Mg}{2} systems requires the detection of the
$\lambda2796$ line and at least one  additional line, the most
convenient being the $\lambda2803$ doublet partner.   We imposed a
5\,$\sigma$  significance level requirement for all $\lambda2796$
lines.  In addition, we also imposed a 3\,$\sigma$ significance level
requirement  for the corresponding $\lambda2803$ line, in order to
ensure identification of the doublet.  In order  to avoid a selection
bias in the $W_0^{\lambda2796} / W_0^{\lambda2803}$ doublet ratio
($DR$), which has a maximum value of $DR=2.0$,  we defined a detection
limit for $W_0^{\lambda2796}$, which we call $W_0^{lim}$, by taking
the larger of 5.0 times the error in $W_0^{\lambda2796}$ or 6.0 times
the error in $W_0^{\lambda2803}$.  The errors were computed using the
unresolved profile width at the given redshift to avoid selection bias
in the Gaussian fit-width.  Only systems 3,000 km s$^{-1}$  blueward
of $z_{QSO}$ and redward of Ly$\alpha$ emission were accepted.
Systems with separations less than 500 km\,s$^{-1}$ were considered
single systems.  This value was chosen to match the width of the
broadest systems found, and is consistent with the window used in the
CRCV99 study.  We also measured \ion{Mg}{1} $W_0^{\lambda2852}$ and
\ion{Fe}{2} $W_0^{\lambda2600}$ for confirmed systems, when possible.

\subsection{Monte Carlo Simulations}
\label{sec_mc}
We ran Monte Carlo simulations of the absorber catalog to test the
efficiency  of our detection technique and to identify biases and
systematic effects.   For each detected system, a simulated doublet
having the same redshift, and similar  $W_0^{\lambda2796}$, $DR$, and
fit width was put into many randomly selected EDR spectra.  Each
spectrum containing a simulated doublet was then run through the
entire non-interactive  and interactive pipelines, and $z$ and
$W_0^{\lambda2796}$ were measured for detected systems.  Thus, we had
two simulated catalogs: the input catalog containing over 9100
simulated doublets, although only $\approx 4500$ of these appear in
regions of spectra with sufficient signal to noise ratio to meet the
detection threshold for the input $W_0^{\lambda2796}$, and the
simulated output catalog, containing $\approx 4400$ doublets recovered
from the  simulation.  Measurement error in $W_0^{\lambda2796}$ and
any subtle systematic  effects can cause systems with
$W_0^{\lambda2796} < W_0^{lim}$ to scatter into the output  catalog as
well as systems with $W_0^{\lambda2796} \ge W_0^{lim}$ to scatter out.
However, if the input distribution of $W_0^{\lambda2796}$ does not
match the true distribution, the ratio of the $\approx 4400$ recovered
systems to  the  $\approx 4500$ input with $W_0^{\lambda2796} \ge
W_0^{lim}$ is not necessarily an accurate estimate of the biases.
Therefore, a trial $W_0^{\lambda2796}$ distribution must be specified
according to which lines are chosen from the simulated input catalog.
The  trial distribution must then be adjusted so that the simulated
output best  represents the data.

As the form of the input distribution is necessarily motivated by the
data, further discussion  of the simulations and their use for
estimating systematic errors are discussed in \S\ref{sec_re}.

\section{Results}
\label{sec_re}
We were able to identify and measure 1331 \ion{Mg}{2} doublets with
$W_0^{\lambda2796} \ge 0.3$\,\AA.  A small number of additional
possible doublets were also discovered ($\lesssim 1\%$ of the sample),
but only allowed  limits on $W_0^{\lambda2796}$  to be determined or
had questionable identifications and were not used in our analyses.

\subsection{$W_0^{\lambda2796}$ Distribution}
\begin{figure}
\epsscale{0.7}
\plotone{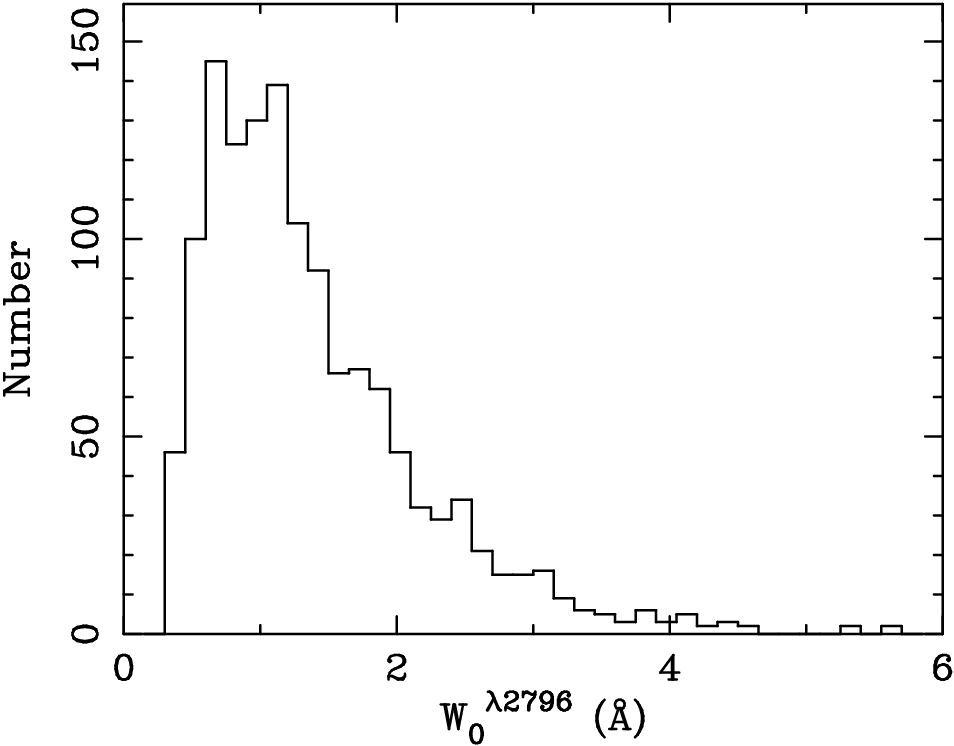}
\caption{The distribution of rest equivalent widths,
$W_0^{\lambda2796}$,  for \ion{Mg}{2} systems found in the survey with
$W_0^{\lambda2796} \ge 0.3$\,\AA.  }
\label{fig_wdist}
\end{figure}

Figure \ref{fig_wdist} shows the distribution of $W_0^{\lambda2796}$
for our sample.  Only those systems with $W_0^{\lambda2796} \ge
0.3$\,\AA\, which we use in our analyses, are shown.  The distribution
has a smooth tail out to  $W_0^{\lambda2796} \approx 4.6$\,\AA, with
the largest value $W_0^{\lambda2796} = 5.68$\,\AA.

\begin{figure}
\plotone{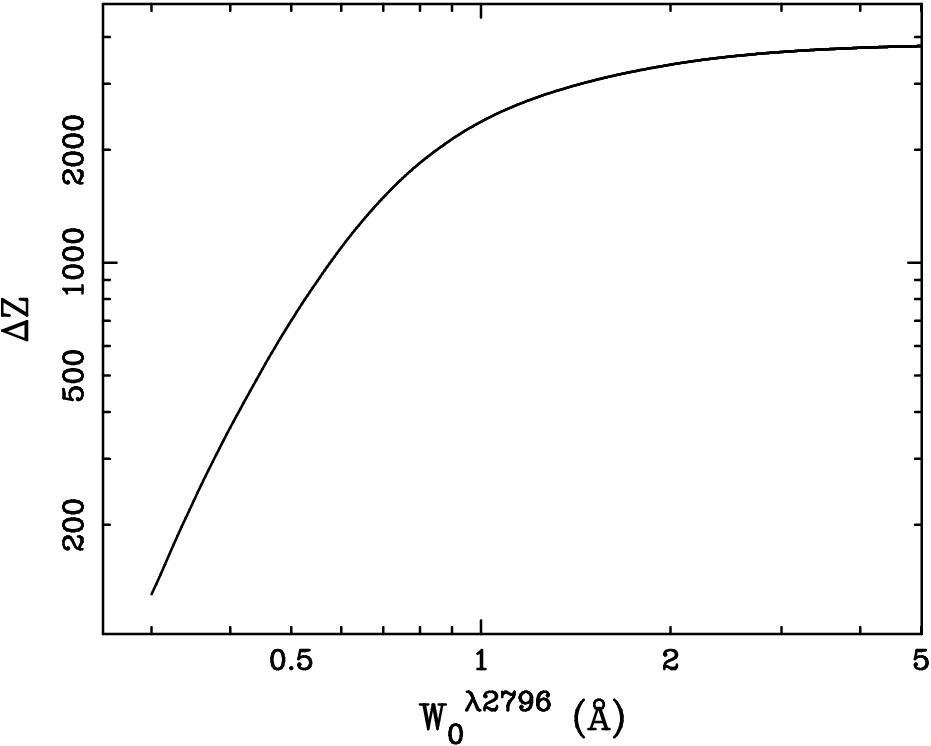}
\caption{The redshift-path covered by the survey, $\Delta
Z\,(W_0^{\lambda2796}) =
\int_{z_{min}}^{z_{max}}\,\sum_i^{N_{spec}}\,g_i(W_0^{\lambda2796},z)\,dz$,
as a function of $W_0^{\lambda2796}$.}
\label{fig_path_rew}
\end{figure}

\begin{figure}
\plotone{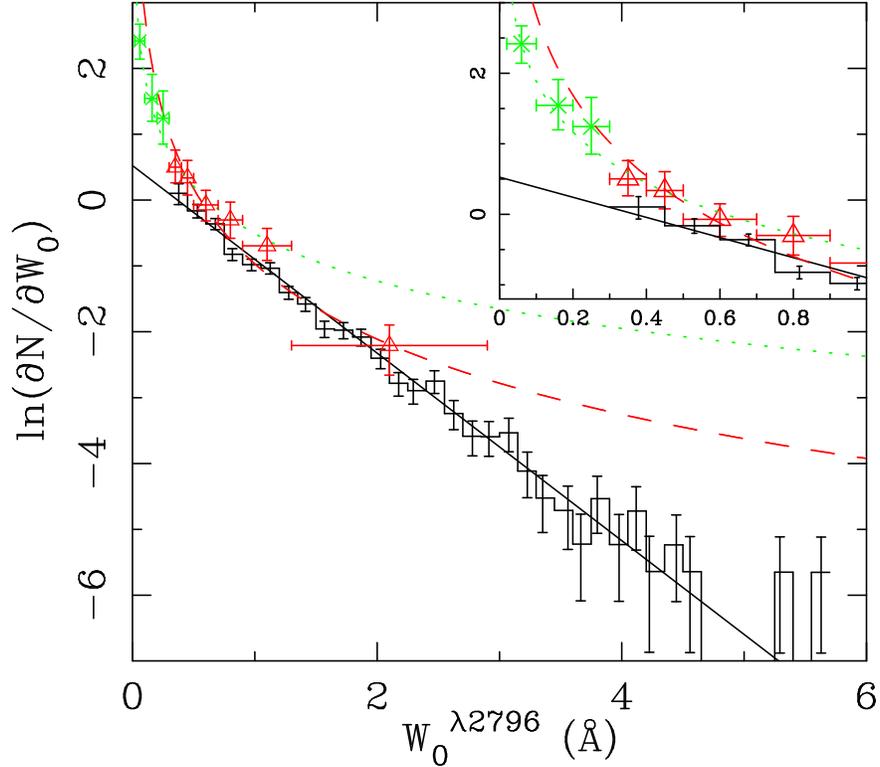}
\caption{Histogram of the $W_0^{\lambda2796}$ distribution.  The solid
line is a maximum likelihood fit of the form 
$\partial N/\partial W_0^{\lambda2796}=
\frac{N^*}{W^*} e^{-\frac{W}{W^*}}$ with $W^*=0.702 \pm 0.017$\,\AA\ and
$N^*=1.187 \pm 0.052$.  The open triangles are from SS92.  The dashed
line is their best fit power law.  The $\times$ symbols represent data
from CRCV99.  The dotted line is their power law fit to their binned
data plus  the SS92 data, excluding the highest-$W_0^{\lambda2796}$
SS92 bin.  The inset shows the $W_0^{\lambda2796} < 1.0$\,\AA\ region in
more detail.  The power laws greatly  over-predict the incidence of
strong ($W_0^{\lambda2796} > 2$\,\AA)  systems, while our exponential
under-predicts the incidence of $W_0^{\lambda2796} < 0.3$\,\AA\ systems.
This suggests a transition in the $\partial N/\partial W_0^{\lambda2796}$ 
distribution
around $W_0^{\lambda2796} \approx 0.3$\,\AA, possibly indicative of
multiple distinct populations.}
\label{fig_rewd}
\end{figure}

The total redshift path covered by our sample for each value of
$W_0^{\lambda2796}$ is given by:
\begin{equation}
\Delta Z\,(W_0^{\lambda2796}) = \int_{z_{min}}^{z_{max}}\,g(W_0^{\lambda2796},z)\,dz =
\int_{z_{min}}^{z_{max}}\,\sum_i^{N_{spec}}\,g_i(W_0^{\lambda2796},z)\,dz,
\end{equation}
where $g_i(W_0^{\lambda2796},z)=1$ if $W_0^{lim}(z) \le
W_0^{\lambda2796}$ and $g_i(W_0^{\lambda2796},z)=0$ otherwise, and the
redshift limits are defined to be 3,000 km s$^{-1}$ above Ly$\alpha$
emission and 3,000 km s$^{-1}$ below \ion{Mg}{2} emission, or the
limits of the data.  BAL regions were masked out.  The redshift path
coverage is  shown in Figure \ref{fig_path_rew}.  Figures
\ref{fig_wdist} and \ref{fig_path_rew}  were combined to form an
unbiased $W_0^{\lambda2796}$  distribution, which is shown in Figure
\ref{fig_rewd}.  The distribution is fit very well by an exponential,
\begin{equation}
\partial N/\partial W_0^{\lambda2796} = \frac{N^*}{W^*}
e^{-\frac{W_0}{W^*}},
\end{equation}
with the maximum likelihood value $W^*=0.702 \pm 0.017$\,\AA\ and
corresponding $N^*=1.187 \pm 0.052$.   The reduced $\chi^2$ comparing
the maximum likelihood fit to the binned data is close to unity,
independent of the choice of bin size.  Also shown are data from
CRCV99 with $W_0^{\lambda2796} < 0.3$\,\AA\ and SS92 with   0.3\,\AA\
$\le W_0^{\lambda2796} \le 2.9$\,\AA.  All three data sets have
similar average absorption redshifts: the CRCV99 data has
$\left<z_{abs}\right> = 0.9$, and our SDSS EDR and the SS92 data have
$\left<z_{abs}\right> = 1.1$.  The SS92 best-fit exponential closely
agrees with our SDSS results,  although our normalization, $N^*/W^*$,
is slightly ($\approx 1\sigma$)  lower than the SS92 value ($W^*=0.66
\pm 0.11$\,\AA, $N^*=1.55 \pm 0.20$).   The reason for this offset, as
well as the lack of scatter as compared to the size of the error  bars
in the SS92 data, is not clear.  We note that both our survey and the
CRCV99 study consider systems within 500 km s$^{-1}$ as a single
system, while the SS92 study uses a 1,000 km s$^{-1}$ window.  This
cannot be the source of the normalization offset, however, as it would
have the opposite effect, reducing the number of systems found by SS92.

A power law fit of the form $\partial N/\partial W_0^{\lambda2796} =
 C\,W_0^{-\delta}$ with  the SS92 values of $C = 0.38$ and $\delta =
 1.65$ is shown as a long-dash line in Figure \ref{fig_rewd}.  It is a
 good fit to our data for  0.5\,\AA\ $\lesssim W_0^{\lambda2796}
 \lesssim 2.0$\,\AA.  For large values of $W_0^{\lambda2796}$,
 however, the SS92 power law over-predicts $\partial N/\partial
 W_0^{\lambda2796}$  by almost an order of magnitude.  CRCV99 also fit
 a  power law to their binned data combined with the SS92 binned data,
 but excluding the  highest SS92 bin.  This is shown as a dotted line.
 The combined data sets suggest  a transition in $\partial N/\partial
 W_0^{\lambda2796}$ occurring near $\approx$ 0.3\,\AA.

We used our simulated catalogs (\S\ref{sec_mc}) to test for biases in
the  $W_0^{\lambda2796}$ distribution.  We chose lines randomly from
the input catalog according to a distribution  of the form $\partial
N/\partial W_0^{\lambda2796} \propto e^{-\frac{W_0}{W^*_{in}}}$, with
an initial guess for $W^*_{in}$, until the number of lines recovered
were equal to that  of the actual catalog, which determines
$N^*_{in}$.  Lines with input $W_0^{\lambda2796} < 0.3$\,\AA\ were
used, although only lines with recovered $W_0^{\lambda2796} \ge
0.3$\,\AA\ were retained.   We determined a maximum-likelihood value
for $W^*_{out}$ using the recovered  $W_0^{\lambda2796}$ values.  We
then corrected our guess for $W^*_{in}$ to minimize $|W^*_{out} -
W^*_{data}|$.  The process was repeated several times with different
seeds of the  random number generator to determine variance.

\begin{figure}
\plotone{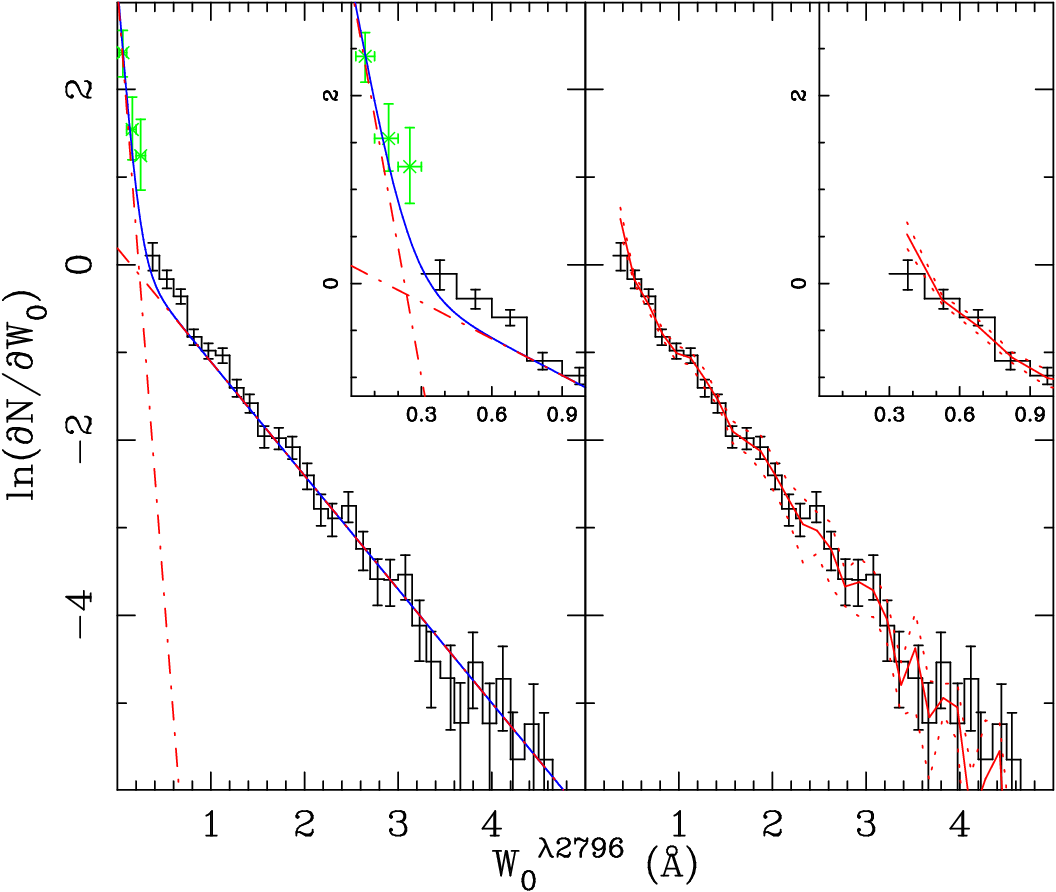}
\caption{Monte Carlo results.  The histograms represent the data.  The
points are from CRCV99.   The left panel shows the input exponential
distributions.   The shallower exponential ($W^*=0.771$\,\AA\ and
$N^*=0.932$) was chosen such that the simulated output distribution
had $W^*_{out} = W^*_{data}$ for $W_0^{\lambda2796} \ge 0.5$\,\AA.
The steeper exponential ($W^*=0.072$\,\AA\ and $N^*=1.71$) was then
chosen to minimize the $\chi^2$  computed by comparing the total
simulated binned data to the real binned data.  The solid line
is the sum of the two exponential fits.  The right panel shows
the simulated output (solid line) over-plotted on the data.  The
dotted lines represent the square root of the variance from different
random number seeds.}
\label{fig_monte}
\end{figure}

Except for the weakest systems in our catalog, which were
under-predicted, this simulation  was able to match the actual data
well.  The under-prediction could be a weakness of our simulation, as
we had few systems with $W_0^{\lambda2796} < 0.3$\,\AA\ with which to
model the  low $W_0^{\lambda2796}$ end of the distribution.
Alternatively, the actual distribution could diverge from the
simulated exponential for $W_0^{\lambda2796} < 0.3$\,\AA, as suggested
by the CRCV99 data.  Thus, though we simulated the entire range
$W_0^{\lambda2796} \ge 0.3$\,\AA,  we limited the calculation of
$W^*_{out}$ to $W_0^{\lambda2796} \ge 0.5$\,\AA.   Motivated by the
CRCV99 results, we then added a second exponential to the input
distribution while holding fixed the primary distribution.  We
adjusted the values of $W^*$ and $N^*$  for this second exponential to
minimize a $\chi^2$ statistic calculated by comparing the binned
simulated output to the binned data.
Thus, we model the input distribution with separate ``weak'' and
``strong'' components:
\begin{equation}
\partial N/\partial W_0^{\lambda2796} =
\frac{N^*_{wk}}{W^*_{wk}}\,e^{-W_0/W^*_{wk}} +
\frac{N^*_{str}}{W^*_{str}}\,e^{-W_0/W^*_{str}}.
\label{eqn2}
\end{equation}
The resulting best fit values are $N^*_{wk} = 1.71\pm0.02$ and
$W^*_{wk}=0.072\pm0.001$\,\AA, and $N^*_{str} = 0.932\pm0.011$ and
$W^*_{str}=0.771\pm0.014$\,\AA, where the errors are the square root
of the variances from different choices of random number generator
seed.  These results are shown in Figure \ref{fig_monte}.  The
``weak'' values were very stable under  changes in the seed, producing
small variances.  Since the ``weak'' component was constrained by only
a small region of $W_0^{\lambda2796}$-space in the data, the actual
uncertainties in the parameters are much larger than the variance.

Although most of the systems in our sample are at least partially
saturated, the distribution of DR values (see section \S\ref{sec_dr})
indicates that the typical degree of saturation increases from a
mixture at $W_0^{\lambda2796} \lesssim 1$\,\AA\ to virtually all
systems being highly saturated at $W_0^{\lambda2796} \gtrsim 2$\,\AA.
Thus, we note the possibility that if the column density distribution
(which is {\it not} measurable with SDSS data)  is described by a
power law over the range of line strengths in our sample,
curve-of-growth effects could conceivably cause the corresponding
$\partial N/\partial W_0^{\lambda2796}$ to deviate from a power law at
$W_0^{\lambda2796} \gtrsim 2$\,\AA\ consistent with our results.
Nonetheless, the directly measurable $\partial N/\partial
W_0^{\lambda2796}$ is parameterized very well by a single exponential
for systems with 0.3\,\AA\ $\le W_0^{\lambda2796} \lesssim 5$\,\AA,
and by two exponentials for  $W_0^{\lambda2796} \ge 0.02$\,\AA.

\subsubsection{Redshift Evolution of $N^*$ and $W^*$}
\begin{figure}
\plotone{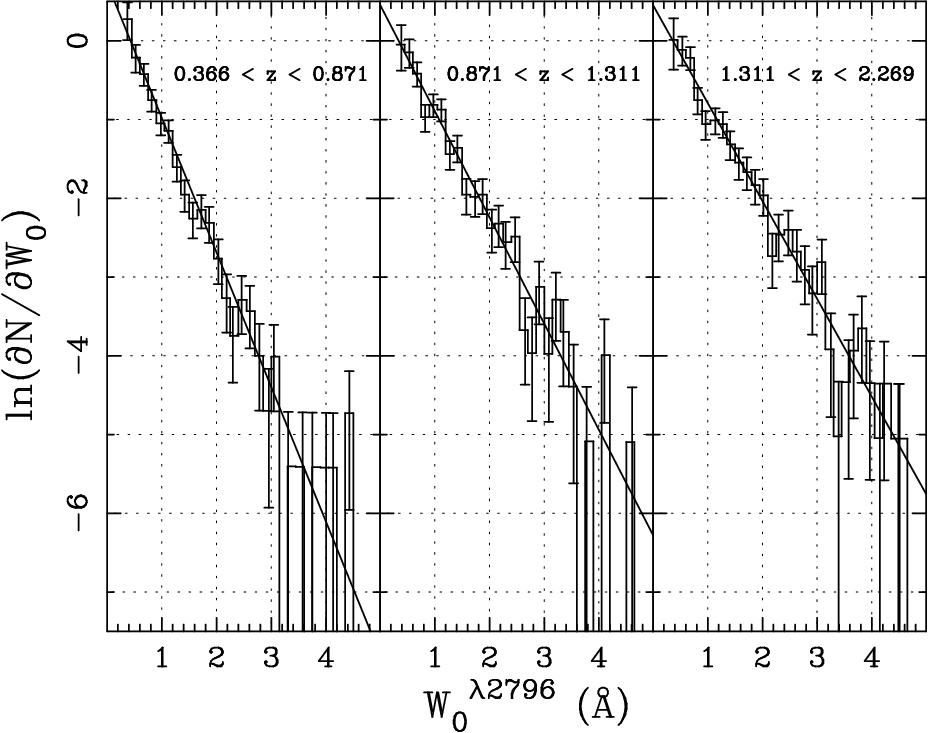}
\caption{The redshift evolution of $\partial N/\partial
W_0^{\lambda2796}$.   The solid lines are maximum likelihood fits of
the form  $\partial N/\partial W_0^{\lambda2796} = \frac{N^*}{W^*}
e^{-\frac{W}{W^*}}$.  Left: $0.366 \le z < 0.871$, with maximum
likelihood value $W^*=0.585 \pm 0.024$\,\AA\ and corresponding
$N^*=1.216 \pm 0.124$.   Center: $0.871 \le z < 1.311$, with maximum
likelihood value $W^*=0.741 \pm 0.032$\,\AA\ and corresponding
$N^*=1.171 \pm 0.083$.  Right: $1.311 \le z < 2.269$, with maximum
likelihood value $W^*=0.804 \pm 0.034$\,\AA\ and corresponding
$N^*=1.267 \pm 0.092$.  Note the steepening of the slope with lower
redshifts.}
\label{fig_rewd_3b}
\end{figure}

\begin{figure}
\plotone{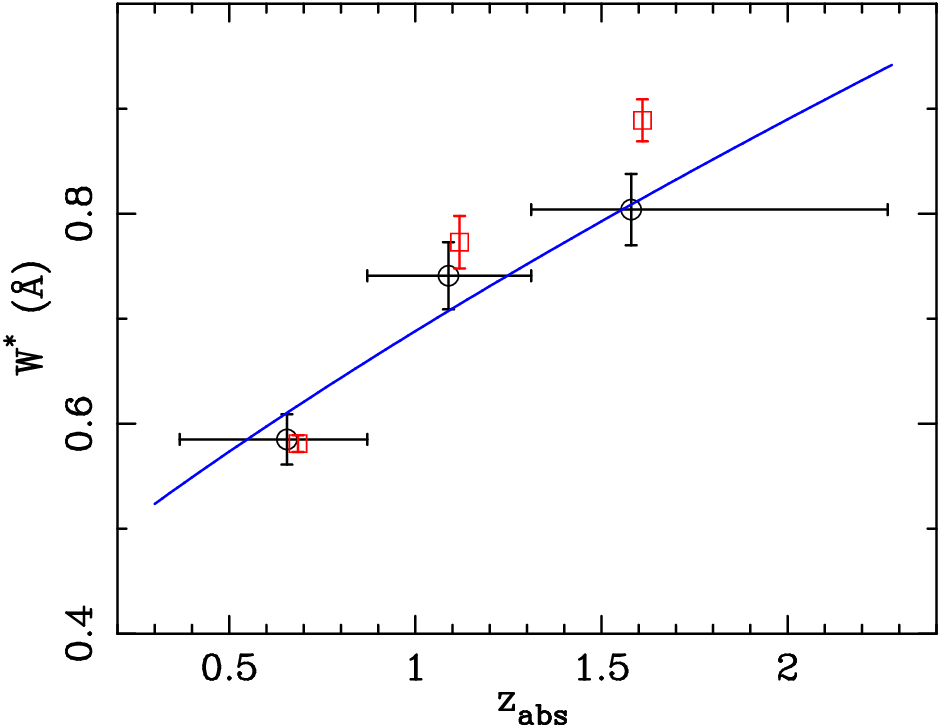}
\caption{The redshift evolution of $W^*$.  The horizontal bars
represent the bin sizes.  The open circles are from the data, while
the squares represent the Monte Carlo input values (see text, \S3.1).
The points are offset slightly in z$_{abs}$ for clarity.  The solid
line is the prediction for  $W^*$ for a fit of the form  $W^*(z) \propto
(1+z)^\beta$ described in \S3.3.}
\label{fig_dwdz}
\end{figure}

In order to investigate evolution in $W^*$, we split our sample into
three redshift bins,  $0.366 \le z < 0.871$, $0.871 \le z < 1.311$,
and $1.311 \le z < 2.269$, chosen  such that there are an equal number
of systems in each bin.  The three $\partial N/\partial
W_0^{\lambda2796}$ distributions are shown in Figure
\ref{fig_rewd_3b}.  Figure \ref{fig_dwdz} shows the resulting $W^*$
for each redshift bin (circles) and the Monte Carlo input values
(squares).  The bars represent the variance from the different random
number seeds only.  The curve in Figure \ref{fig_dwdz} is the power
law fit described in \S\ref{sec_wzd}.

\subsection{Distribution of Absorption Redshifts}
\begin{figure}
\plotone{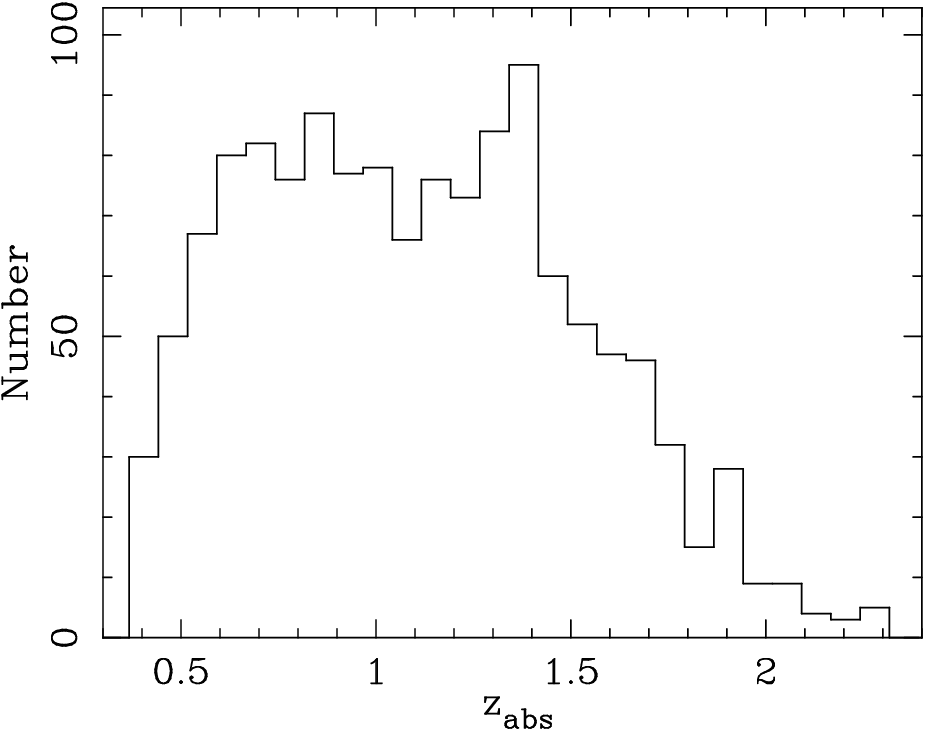}
\caption{The distribution of absorption redshifts for  \ion{Mg}{2}
systems with $W_0^{\lambda2796} \ge 0.3$\,\AA\ found in the survey.}
\label{fig_nz}
\end{figure}

\begin{figure}
\plotone{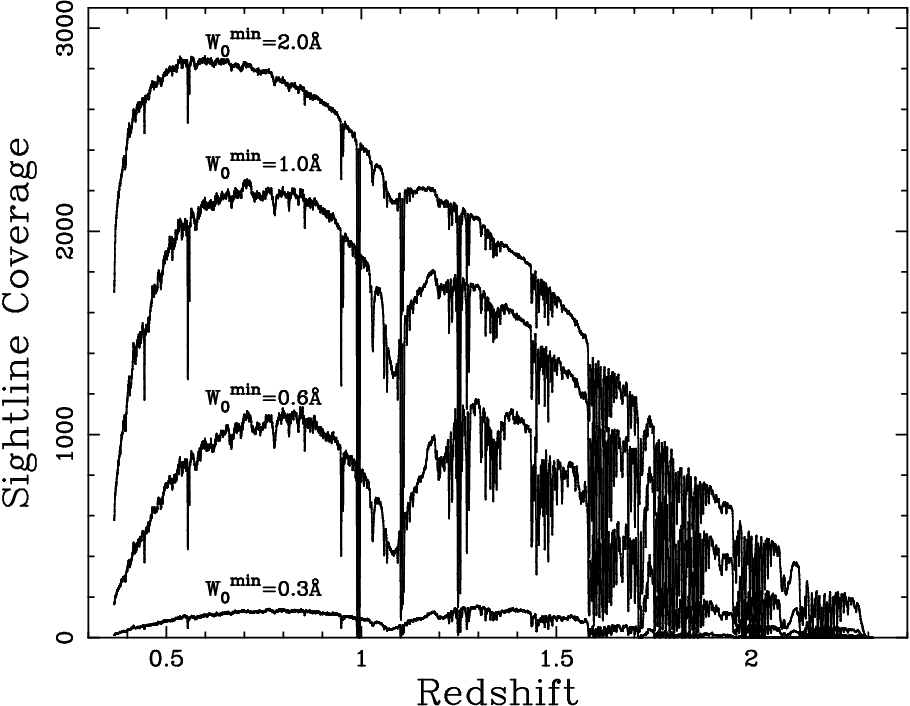}
\caption{The total number of  sightlines with sufficient signal to
noise ratio to detect lines with  $W_0^{\lambda2796} \ge W_0^{min}$ as
a function of redshift,  for $W_0^{min}$ = 2.0\,\AA, 1.0\,\AA,
0.6\,\AA, and 0.3\,\AA. The conspicuous features  at $z > 1.5$ are due
to poor night sky line subtractions in  many of the spectra.  The
depression near $z = 1.1$ is due to the dichroic.}
\label{fig_sl}
\end{figure}

Figure \ref{fig_nz} shows the absorption redshift distribution for our
sample.   Absorption redshifts span  the range $0.367 \le z \le
2.269$.  The total number of sightlines with sufficient  signal to
noise ratio to detect lines with $W_0^{\lambda2796} \ge W_0^{min}$ for
several values of $W_0^{min}$ is shown in Figure \ref{fig_sl} as a
function of redshift.  The conspicuous features at $z > 1.5$ are due
to poor night sky subtractions in  many of the spectra.  The
depression near $z = 1.1$ is due to the dichroic (Stoughton et
al. 2002.)

\begin{figure}
\plotone{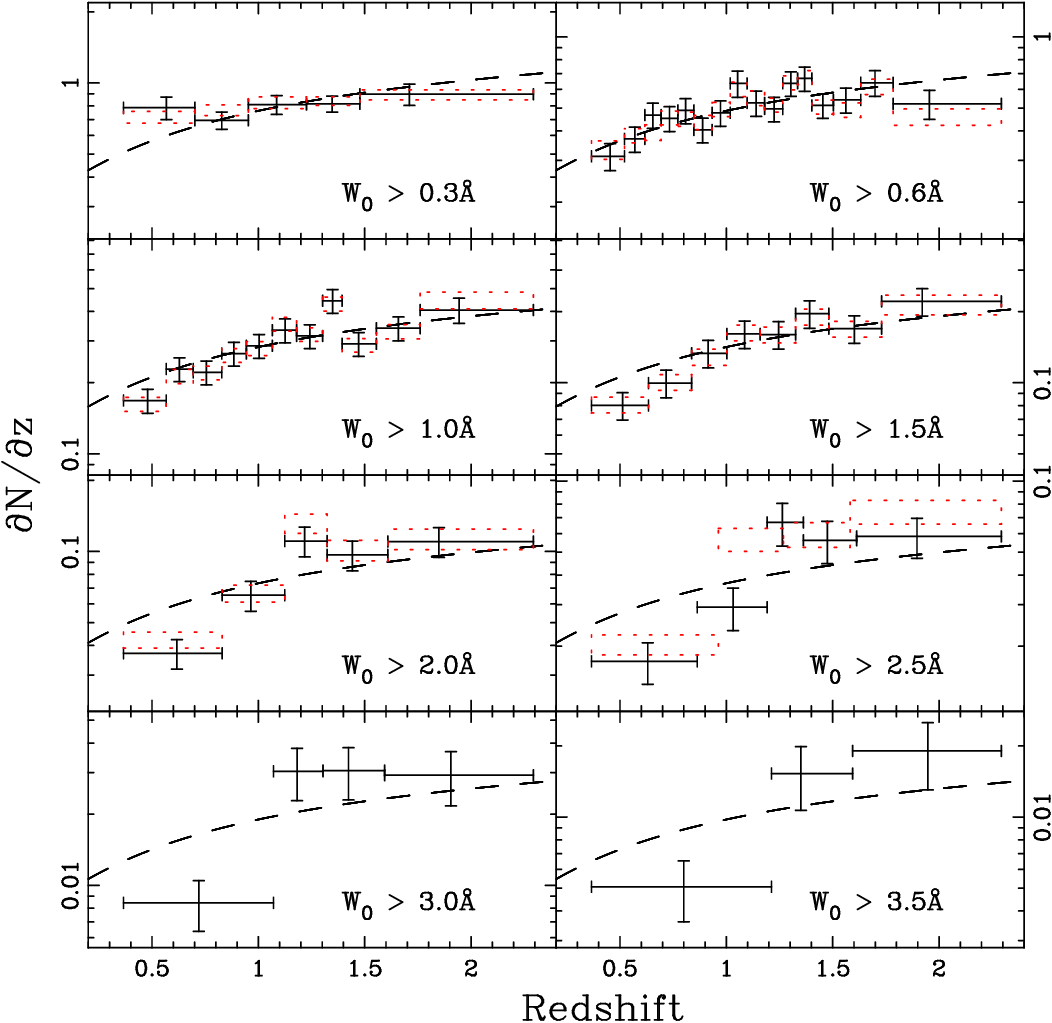}
\caption{Number density evolution of \ion{Mg}{2} absorbers for
$W_0^{\lambda2796} \ge W_0^{min}$ samples.  The dashed lines are the
no-evolution curves for a WMAP cosmology,
($\Omega_M,\Omega_\Lambda,h_0) = (0.3,0.7,0.7)$.  The curves are
normalized to minimize the  $\chi^2$ to the binned data.  The
$W_0^{min}=$ 0.3\,\AA, 0.6\,\AA, 1.0\,\AA, and 1.5\,\AA\ samples have
$\chi^2$ values that are consistent with no evolution.  The
$W_0^{min}=$ 2.0\,\AA,  2.5\,\AA, and 3.0\,\AA\ samples are
inconsistent with the no evolution curves at $\gtrsim3\sigma$, while
the $W_0^{min}=$ 3.5\,\AA\ sample is inconsistent at $\simeq2\sigma$.
The dotted boxes represent the results of the Monte Carlo simulation.
The widths correspond to the bin sizes and the heights to the $\pm
1\sigma$ values.  The $W_0^{min}=$ 3.0\,\AA\ and  3.5\,\AA\ samples
were not large enough to permit meaningful Monte Carlo simulations.}
\label{fig_dndz_cumu}
\end{figure}

The incidence and variance of lines in an interval of
$W_0^{\lambda2796}$ over a specified redshift range are given by:
\begin{equation}
 \frac{\partial N}{\partial z} = \sum_i\frac{1}{\Delta Z\,(W_0^i)},\ \
\ \sigma^2_{\frac{\partial N}{\partial z}} =
\sum_i\,\left(\frac{1}{\Delta Z\,(W_0^i)}\right)^2,
\end{equation}
where the sum is over systems with $W_0^i$ in the given interval and
$\Delta Z\,(W_0^i)$ represents the path contained in the specified
redshift range.  Traditionally, $\partial N/\partial z$ has been
plotted versus redshift for lines stronger than a specified
$W_0^{min}$.  In Figure \ref{fig_dndz_cumu}, we show $\partial
N/\partial z$ as a function of redshift for $W_0^{\lambda2796} \ge$
0.3\,\AA, 0.6\,\AA, 1.0\,\AA, 1.5\,\AA, 2.0\,\AA,  2.5\,\AA, 3.0\,\AA\
and 3.5\,\AA.  Also shown in  Figure \ref{fig_dndz_cumu} are the
no-evolution curves (NECs) for a cosmology with WMAP results (Spergel
et al. 2003), ($\Omega_M,\Omega_\Lambda,h) = (0.3,0.7,0.7)$, scaled to
minimize the $\chi^2$ to the  binned data.  The $W_0^{min}=$ 0.3\,\AA,
0.6\,\AA, 1.0\,\AA, and 1.5\,\AA\ samples have $\chi^2$ values that
are consistent with no evolution.  The $W_0^{min}=$ 2.0\,\AA,
2.5\,\AA, and 3.0\,\AA\ samples are inconsistent with the NECs at
$\gtrsim3\sigma$, while the $W_0^{min}=$ 3.5\,\AA\ sample is
inconsistent at $\approx2\sigma$.  The dotted-boxes in Figure
\ref{fig_dndz_cumu}  show the results of the Monte Carlo simulation
described in \S\ref{sec_mc}.

\begin{figure}
\plotone{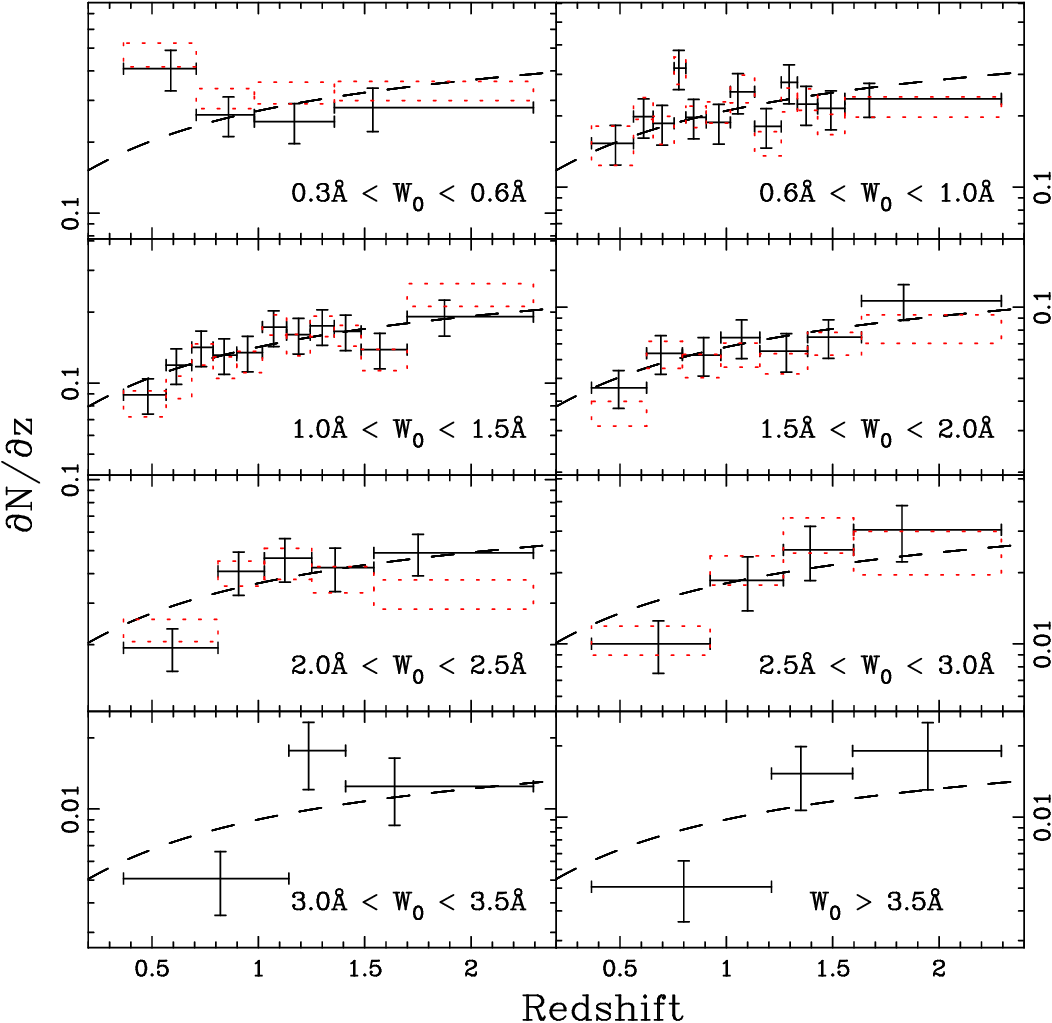}
\caption{Number density evolution of \ion{Mg}{2} absorbers for ranges
of $W_0^{\lambda2796}$.   The dashed lines are the no-evolution curves
for a WMAP cosmology,  ($\Omega_M,\Omega_\Lambda,h_0) =
(0.3,0.7,0.7)$.  The curves are normalized to minimize the  $\chi^2$
to the binned data.  The  0.6\,\AA\ - 1.0\,\AA, 1.0\,\AA\ - 1.5\,\AA,
1.5\,\AA\ - 2.0\,\AA, 2.0\,\AA\ - 2.5\,\AA\ and 2.5\,\AA\ - 3.0\,\AA\
samples  have $\chi^2$ values that are consistent with no evolution.
The no-evolution curve for the 0.3\,\AA\ - 0.6\,\AA\ sample is ruled
out at $\simeq2.5\sigma$ and for  the 3.0\,\AA\ - 3.5\,\AA\ and
$\ge3.5$\,\AA\ samples at $\simeq2.0\sigma$.  The dotted boxes
represent the  results of the Monte Carlo simulation.  The widths
correspond to the bin sizes and the heights to the $\pm 1\sigma$
values.  The 3.0\,\AA\ - 3.5\,\AA\ and greater than 3.5\,\AA\ samples
were not large enough to permit meaningful Monte Carlo simulations.}
\label{fig_dndz_nec}
\end{figure}

The large size of our data set allows us to investigate $\partial
N/\partial z$ not only for distributions cumulative in
$W_0^{\lambda2796}$, but also for ranges of $W_0^{\lambda2796}$.  This
is potentially important, as evolution in the largest
$W_0^{\lambda2796}$ values  is not necessarily negligible in
cumulative $\partial N/\partial z$ distributions.   Thus, we repeated
the above analysis for the following ranges: 0.3\,\AA\ $\le
W_0^{\lambda2796} <$ 0.6\,\AA, 0.6\,\AA\ $\le W_0^{\lambda2796} <$
1.0\,\AA, 1.0\,\AA\ $\le W_0^{\lambda2796} <$ 1.5\,\AA, 1.5\,\AA\ $\le
W_0^{\lambda2796} <$ 2.0\,\AA, 2.0\,\AA\ $\le W_0^{\lambda2796} <$
2.5\,\AA, 2.5\,\AA\ $\le W_0^{\lambda2796} <$ 3.0\,\AA, 3.0\,\AA\ $\le
W_0^{\lambda2796} <$ 3.5\,\AA, and $W_0^{\lambda2796} \ge$ 3.5\,\AA.
The results  are shown in Figure \ref{fig_dndz_nec}.  The 0.6\,\AA\ -
1.0\,\AA, 1.0\,\AA\ - 1.5\,\AA, 1.5\,\AA\ - 2.0\,\AA, 2.0\,\AA\ -
2.5\,\AA\ and 2.5\,\AA\ - 3.0\,\AA\ samples have $\chi^2$ values that
are consistent with no evolution.   The NEC for the 0.3\,\AA\ -
0.6\,\AA\ sample is ruled out at $\approx2.5\sigma$ and for the
3.0\,\AA\ - 3.5\,\AA\ and $\ge$ 3.5\,\AA\ samples at
$\approx2.0\sigma$.

\begin{figure}
\plotone{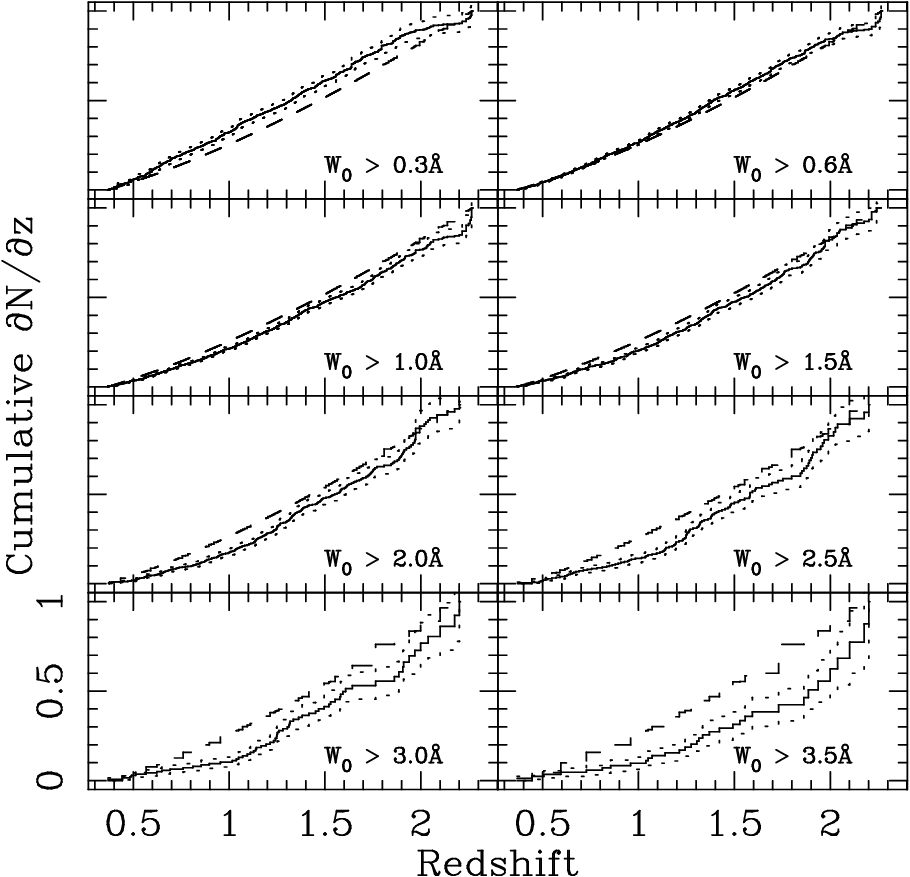}
\caption{The cumulative distribution of $\partial N/\partial z$ (solid
lines) for the $W_0^{\lambda2796} \ge W_0^{min}$ samples, compared
with the cumulative  no-evolution curves (dashed lines) for a WMAP
cosmology ($\Omega_M,\Omega_\Lambda,h_0) = (0.3,0.7,0.7)$.  The dotted
curves represent the $1\sigma$ levels.  The no evolution curves
under-predict $\partial N/\partial z$ at low redshift for the
$W_0^{min}=$ 0.3\,\AA\ sample,  and over-predict $\partial N/\partial
z$ at low redshift for the larger $W_0^{min}$ samples.   The data are
inconsistent with the NECs at more than $3\sigma$ for all but the
$W_0^{min}=$ 0.6\,\AA\ sample.}
\label{fig_ks1}
\end{figure}

\begin{figure}
\plotone{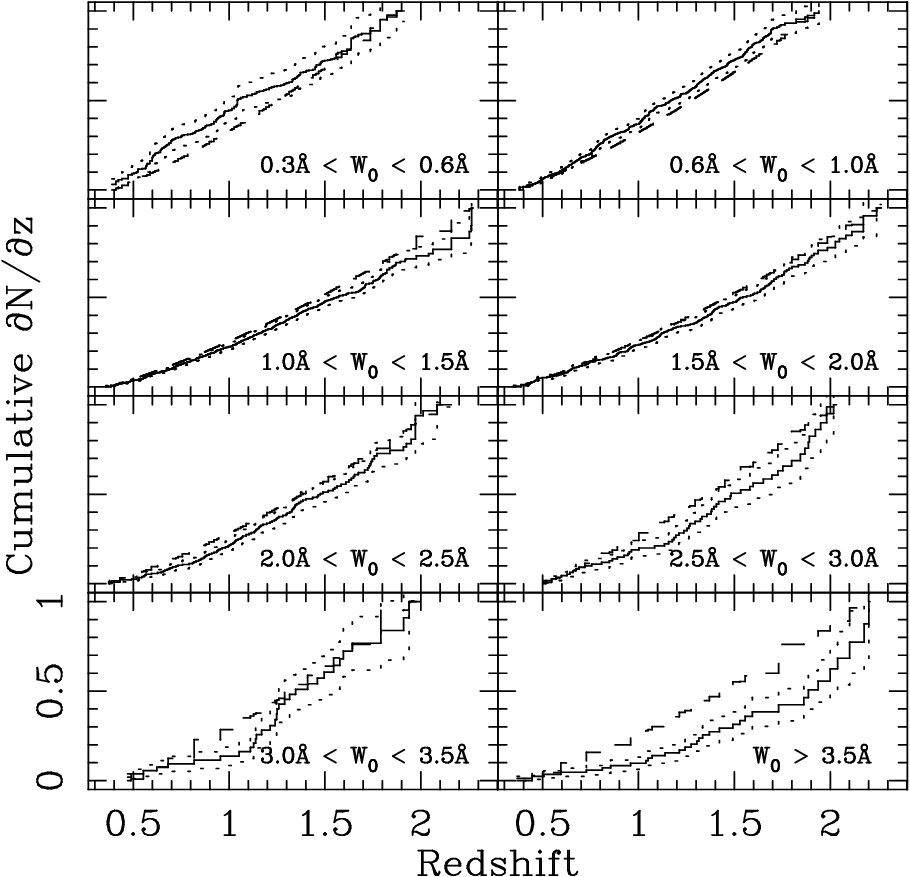}
\caption{The cumulative distribution of $\partial N/\partial z$ (solid
lines) for the $W_0^{\lambda2796}$ ranges, compared with the
cumulative  no-evolution curves (dashed lines) for a WMAP cosmology
($\Omega_M,\Omega_\Lambda,h_0) = (0.3,0.7,0.7)$.  The dotted curves
represent the $1\sigma$ levels.  The no evolution curve under-predicts
$\partial N/\partial z$ at low redshift for the $W_0^{\lambda2796}=$
0.3\,\AA\ - 0.6\,\AA\ and $W_0^{\lambda2796}=$ 0.6\,\AA\ - 1.0\,\AA\
samples, and over-predicts $\partial N/\partial z$ at low redshift for
the larger $W_0^{\lambda2796}$ ranges.  However, the data rule out the
NECs at $\ge3\sigma$ only for the $W_0^{\lambda2796} \ge 2.0$\,\AA\
samples.}
\label{fig_ks2}
\end{figure}

Since the NEC normalization is a free parameter, a plot cumulative in
redshift comparing $\partial N/\partial z$ from the data to the NEC is
more instructive.   These plots, shown in Figures \ref{fig_ks1} and
\ref{fig_ks2}, highlight the skew of the $\partial N/\partial z$
curves from the NEC prediction that is not necessarily manifested in
the $\chi^2$ analysis.   The NECs over-predict $\partial N/\partial z$
at low redshift for small $W_0^{min}$ and under-predict $\partial
N/\partial z$ at low redshift for large $W_0^{min}$.  The transition
occurs for the  cumulative $W_0^{\lambda2796}$ plots between
$W_0^{min} =$ 0.6\,\AA\ and 1.0\,\AA\, and the effect is stronger for
increasingly larger values of $W_0^{min}$.  The plots using ranges of
$W_0^{\lambda2796}$ show that the transition from over- to under-
predicting $\partial N/\partial z$ at lower redshift occurs around
1\,\AA, and significant detection  of evolution is seen in systems
with $W_0^{\lambda2796} \gtrsim$ 2.0\,\AA.  The evolution signal is
strong ($\approx 4\sigma$) for lines with $W_0^{\lambda2796} \ge$
3.5\,\AA.  Although these plots do give a clearer indication of the
deviation from the NECs, it is difficult  to ascribe a K-S probability
to the curves in Figures \ref{fig_ks1} and \ref{fig_ks2} because
systems contribute to $\partial N/\partial z$ in a non-uniform manner,
i.e., inversely proportional to $g(W_0,z)$.

The $\partial N/\partial z$ point in the lowest redshift bin for the
$W_0^{\lambda2796} \ge$ 0.3\,\AA\  sample in Figure
\ref{fig_dndz_cumu}, and for the 0.3\,\AA\ $\le W_0^{\lambda2796} <$
0.6\,\AA\ sample in  Figure \ref{fig_dndz_nec}, lie well above the
NEC.  The  increase is greater in magnitude but smaller in
significance ($\approx2\sigma$ versus $\approx3\sigma$) for the
non-cumulative sample.  The Monte Carlo results lessen the
significance in the cumulative sample, but not in the non-cumulative
sample.  Thus, the weakest lines in our  study show evolution in the
sense that their incidence {\it increases} with decreasing redshift.
However, this may be an artifact of the sharp cutoff in our sample at
0.3\,\AA.  A sample with $W_0^{\lambda2796} \ge$ 0.4\,\AA\ is
consistent with the NEC,  and $\partial N/\partial z$ for 0.4\,\AA\
$\le W_0^{\lambda2796} <$ 0.6\,\AA\ is within $1\sigma$ of the NEC at
all redshifts.   We cannot exclude the possibility, however, that the
increase is real.

A comparison of the $\partial N/\partial z$ values from our survey and 
SS92 for the full redshift 
ranges are shown in Table 1.  Our $W_0^{\lambda2796} \ge$ 0.6\,\AA\
and 1.0\,\AA\ results are consistent  with SS92.  However, we find a
smaller value of $\partial N/\partial z$ for $W_0^{\lambda2796} \ge$
0.3\,\AA\ (although this difference is confined to $z \gtrsim 1.3$.)

\subsection{Joint $W_0^{\lambda2796}$-Redshift Distribution}
\label{sec_wzd}
Conventionally, the $W_0$-redshift distribution of absorption lines,
$\partial ^2N/\partial z\,\partial W_0$, has been parameterized by a
combination of a power law in redshift and exponential in $W_0$ (e.g.,
Murdoch et al. 1986; Lanzetta, Turnshek \& Wolfe 1987; Weymann et
al. 1998):
\begin{equation}
dN/dW_0 = \frac{N^*}{W^*}e^{-\frac{W_0}{W^*}}, \ \
dN/dz = N_0\,(1+z)^\gamma.
\end{equation}
This parameterization is convenient for the determination of best-fit
values for the  parameters because of the separate $W_0$ and $z$
dependences.  
However, our data reveal that a single power law is not a good fit to 
$\partial N/\partial z$ for any given range of $W_0^{\lambda2796}$.  
Also, we have shown that $W^*$ depends on redshift
for  \ion{Mg}{2} $\lambda2796$ lines (i.e., $\partial N/\partial
z$ for different ranges of $W_0^{\lambda2796}$ varies differently with
redshift.)
We do not wish to abandon the exponential form, however, as
we have also shown that it is a very good parameterization of the data
at all redshifts for the range of $W_0^{\lambda2796}$ covered by our
sample.  Thus, we retain the general form $\partial N/\partial
W_0^{\lambda2796} = \frac{N^*}{W^*}\,e^{-\frac{W_0}{W^*}}$, but allow
both $N^*$ and $W^*$ to vary with redshift as a power law in $(1+z)$.
The maximum likelihood results are $N^* =
1.001\pm0.132\,(1+z)^{0.226\pm0.170}$ and
$W^*=0.443\pm0.032\,(1+z)^{0.634\pm0.097}$\,\AA.  The values are
highly correlated, and the errors include the effects of the
correlations.  See the Appendix for details of our parameterization
and a discussion of the errors.

\begin{figure}
\plottwo{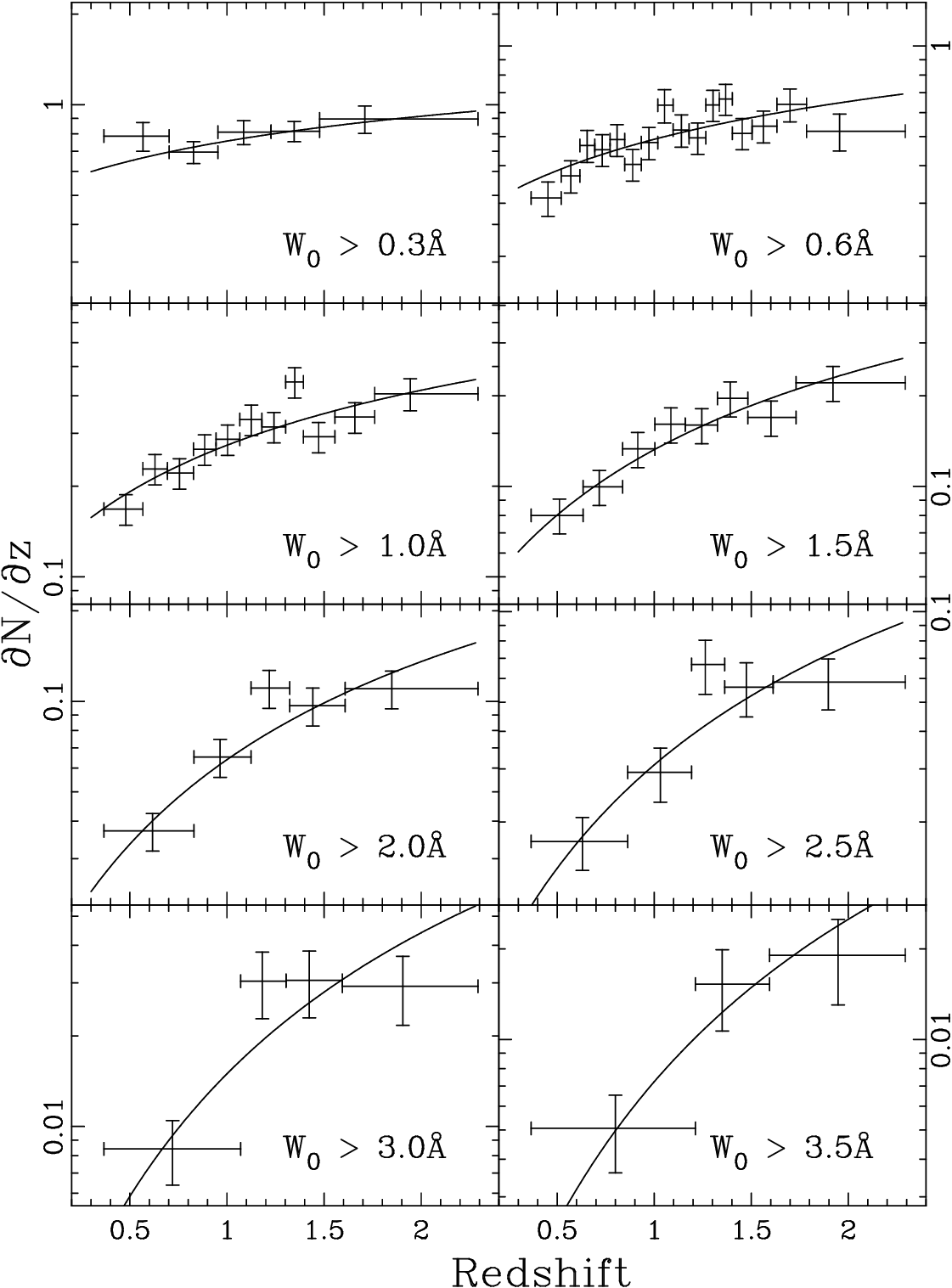}{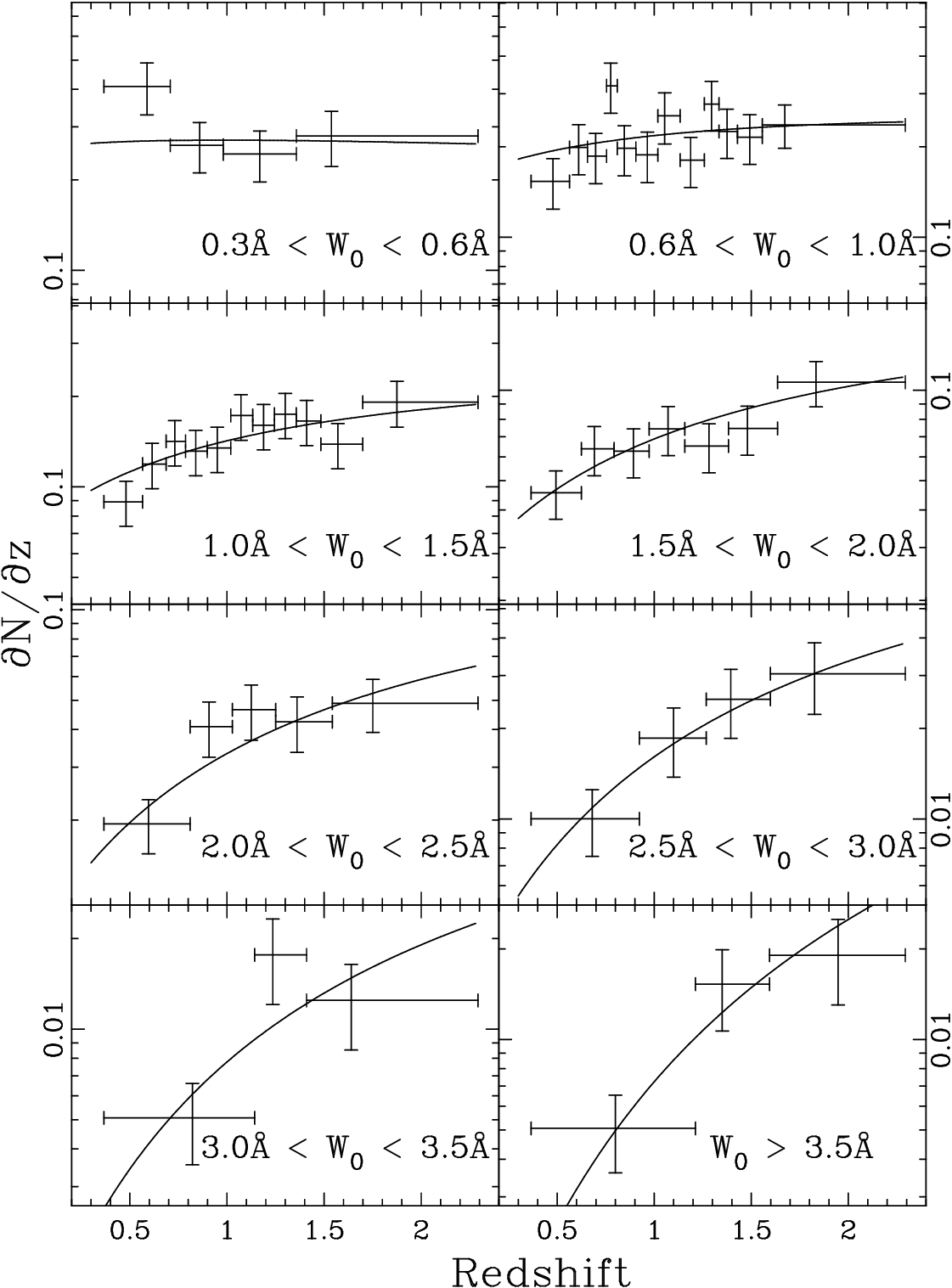}
\caption{$\partial N/\partial z$ curves calculated from our
parameterization (\S\ref{sec_wzd}).  The $\partial N/\partial z$ data
are consistent with the fit for all ranges shown.  Left: The
cumulative samples.  Right: The non-cumulative samples.}
\label{fig_dndzf}
\end{figure}

\begin{figure}
\plottwo{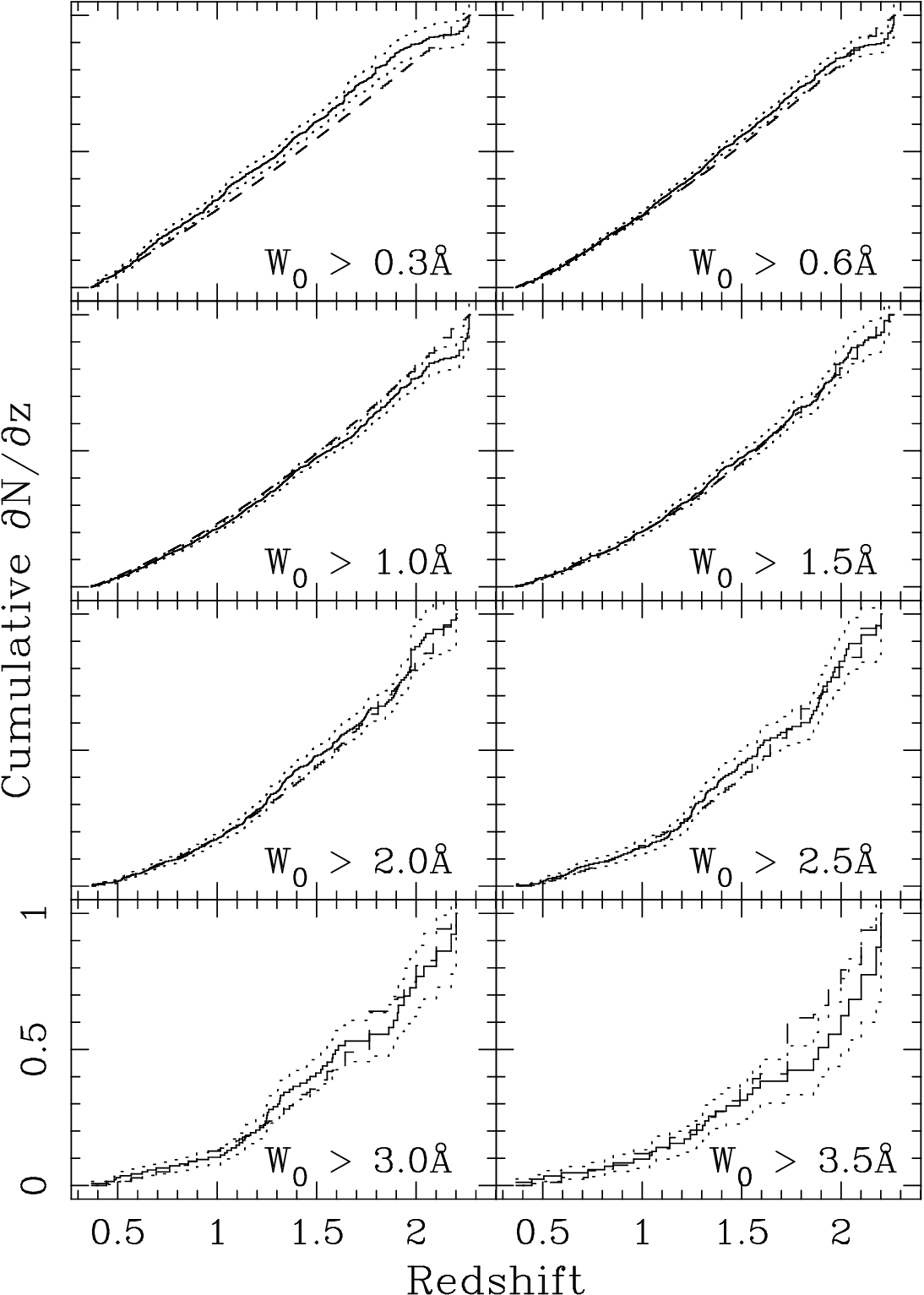}{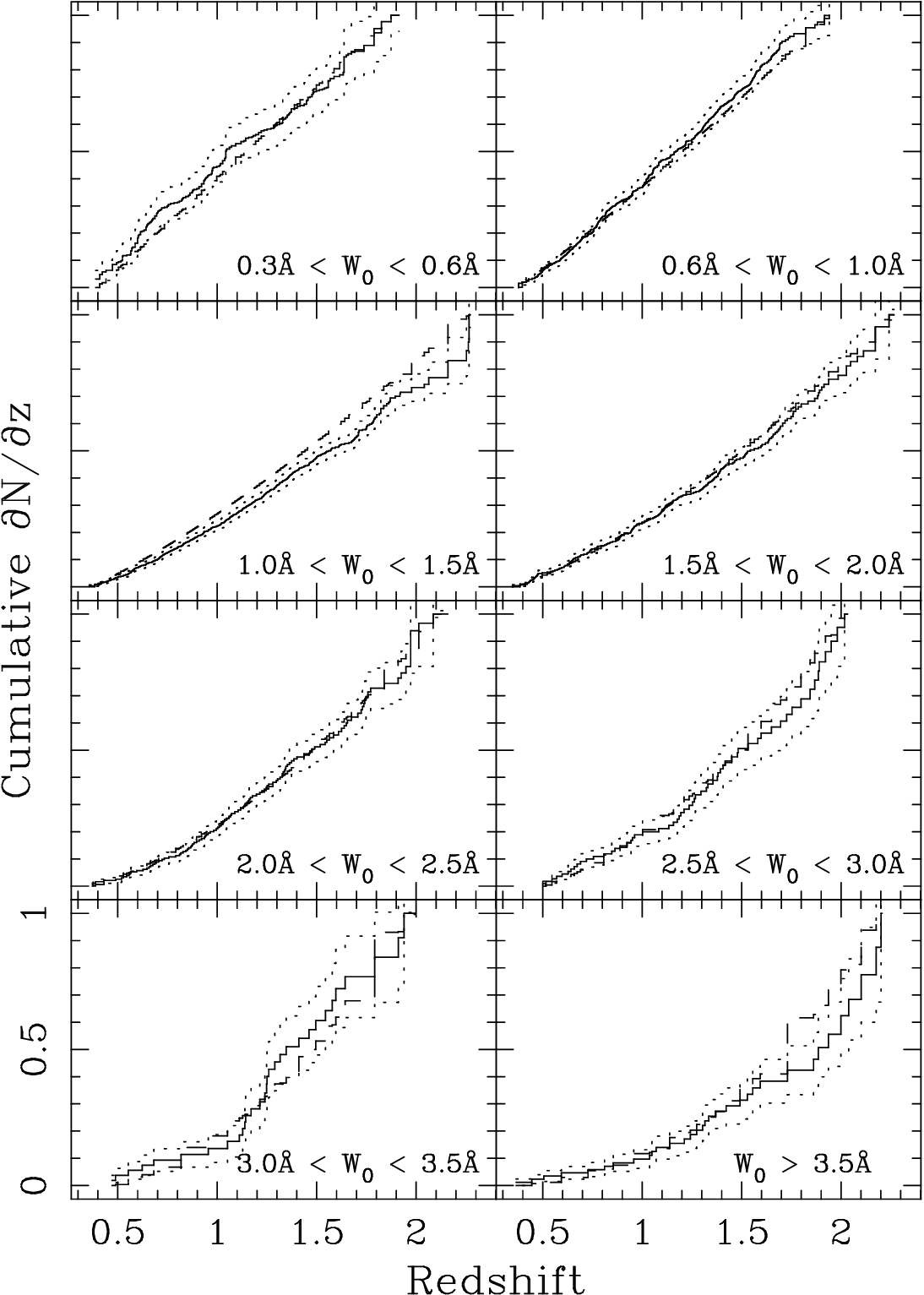}
\caption{The cumulative distribution of $\partial N/\partial z$
plotted against the cumulative  $\partial N/\partial z$ curves
calculated from our parameterization (\S\ref{sec_wzd}).  The $\partial
N/\partial z$ data are consistent with the fit for all ranges shown.}
\label{fig_ks_cs_fit}
\end{figure}

We repeated the analyses of \S3.2, comparing the $\partial N/\partial
z$ curves to those determined using our redshift parameterization of
$N^*$ and $W^*$.   The results, shown in Figures \ref{fig_dndzf} and
\ref{fig_ks_cs_fit},  indicate that our parameterization is indeed a
good description of all of the data.  The evolution is well described
by a steepening of the $W_0^{\lambda2796}$ distribution with a
normalization that changes so as to keep $\partial N/\partial z$
constant for the smaller $W_0^{\lambda2796}$ values.  Extrapolating
our parameterization to lower redshift, we predict $\partial
N/\partial z=0.28 \pm 0.03$ for  $W_0^{\lambda2796} \ge 0.6$\,\AA\ and
$\left<z\right> = 0.06$ and $\partial N/\partial z=0.11 \pm 0.02$ for
$W_0^{\lambda2796} \ge 1.0$\,\AA\ and $\left<z\right> = 0.04$, in
comparison  to Churchill's (2001) values of $\partial N/\partial
z=0.22^{+0.12}_{-0.09}$ and $\partial N/\partial
z=0.16^{+0.09}_{-0.05}$, respectively.  However, our parameterization
is likely inappropriate for describing  $W_0^{\lambda2796} <
0.3$\,\AA\ lines, since not including the apparently two-component
nature of the $W_0^{\lambda2796}$ distribution leads to underestimates
of the number density of weak lines (Figure \ref{fig_rewd}).

Table 2 summarizes the various parameterizations of the data.

\subsection{Systematic Errors}
\label{sec_se}
Our Monte Carlo simulations did not uncover any significant systematic
errors.  All of the Monte Carlo $\partial N/\partial z$ values are
consistent with those from the data.  However, since we used the data
to model the simulated lines, if there are lines with characteristics
that make them underrepresented in the data, they would  also be
underrepresented in the simulated catalogs.  We ran simulations  to
address this issue as we developed our line-finding algorithm.
Although it is possible that we miss certain types of lines, such as
weak kinematic outliers or  lines with abnormally broad or exotic
profiles, for example, these effects are likely to be small.

A potentially more serious source of systematic error may arise in the
regions of poorly subtracted  night sky lines seen in some of the
spectra.  The error arrays are not always accurate in these  regions,
and doublets falling between sky lines can be confused with  residuals
from the  poor subtraction.  As the errors are larger (though not
necessarily accurate) in the  night sky regions, the values of
$W_0^{lim}$ are large as well.  Thus, only the strongest lines  would
be affected.  In principle, our simulations  should account for these
effects.  However, since non-Gaussian profiles are preferentially
found among the stronger lines, their simulation is somewhat less
reliable.  Also, their numbers are much smaller, providing fewer lines
to serve as models in the simulation, and less overall significance.
The largest $W_0^{\lambda2796}$ ranges were, in fact, too sparse for
meaningful simulations.  Errors of this type would be manifest in the
largest $W_0^{\lambda2796}$ ranges and for redshifts $z \sim 1.6 -
2.0$.

Finally, in QSO absorption-line studies it is of interest to asses if
biases due to  the presence of the absorber affecting the magnitude
and color of the background QSO are present.  The most often discussed
effect is that of a dimming and reddening of the background QSO due to
dust in the absorber.  However,  the presence of an  absorbing galaxy
could also contribute light and/or have a lensing effect, causing the
QSO to appear brighter.  These competing effects are investigated,
using the absorbers and simulations from this work,  in M\'{e}nard,
Nestor, \& Turnshek (2004).    They find that QSOs with \ion{Mg}{2}
absorbers tend to be brighter in red passbands and fainter in blue
passbands, indicating a combination of brightening and reddening.  The
effects are stronger for increasingly stronger $W_0^{\lambda2796}$
systems.  The average effects are mostly  within $\pm 0.2$ magnitudes.
As the SDSS EDR QSO catalog selection properties were  not necessarily
homogeneous, it is difficult  to quantify the effect these results may
have on our data.
If these effects are non-negligible, they could affect the measured
$W_0^{\lambda2796}$ distribution and $\partial N/\partial z$.  The
biases are likely to be small, however,  since it would be peculiar if
they conspired to drive $\partial N/\partial W_0^{\lambda2796}$ to
more closely resemble an exponential, and there is no detectable
deviation from an exponential,  consistent with the expected biases,
seen in Figure \ref{fig_rewd}.  Since dust increases with decreasing redshift
(consistent with the findings of Nestor et al.  2003), it is worth
considering if the evolution in $\partial N/\partial z$ seen for the
strongest systems is an affect of increasing dust.  However,
M\'{e}nard et al. find no such increase in bias with decreasing
redshift.  Also, Ellison et al. (2004) compare \ion{Mg}{2} $\partial
N/\partial z$ determined from radio-selected CORALS survey spectra to
values obtained with optical surveys (including this work), and find
good agreement (though they do not investigate the largest
$W_0^{\lambda2796}$ ranges).  Therefore,  the steepening of the
$W_0^{\lambda2796}$ distribution due to a disappearance of the largest
$W_0^{\lambda2796}$ lines at low redshift and the evolution in
$\partial N/\partial z$ for  strong lines appears to be real.

While more recent SDSS QSO spectra  offer the opportunity to increase
the \ion{Mg}{2} absorber sample by another order of magnitude,
systematic errors in line identification and measurement will begin to
dominate in the determination of absorber property statistics.

\subsection{\ion{Mg}{2} Doublet Ratio}
\label{sec_dr}
\begin{figure}
\plottwo{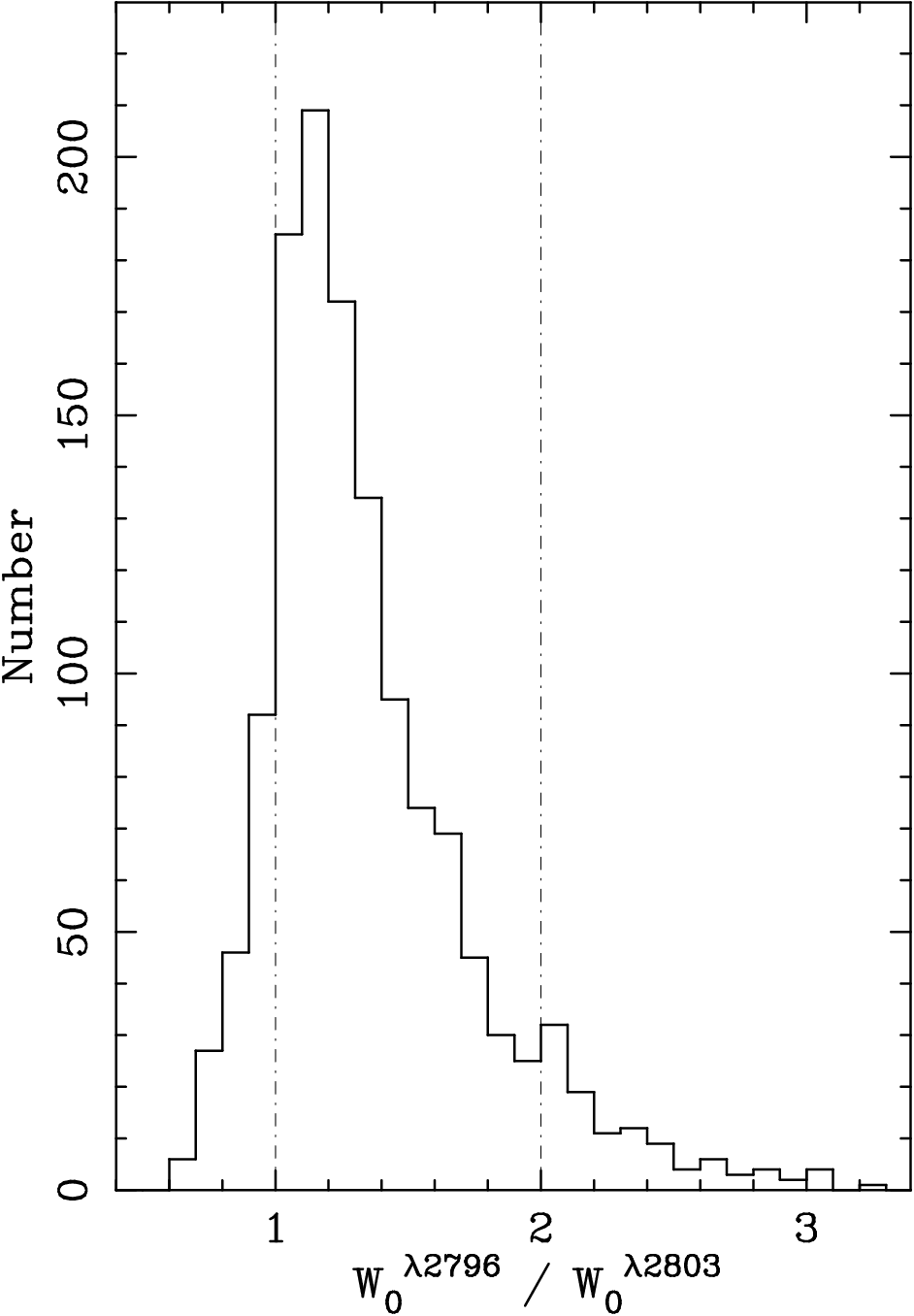}{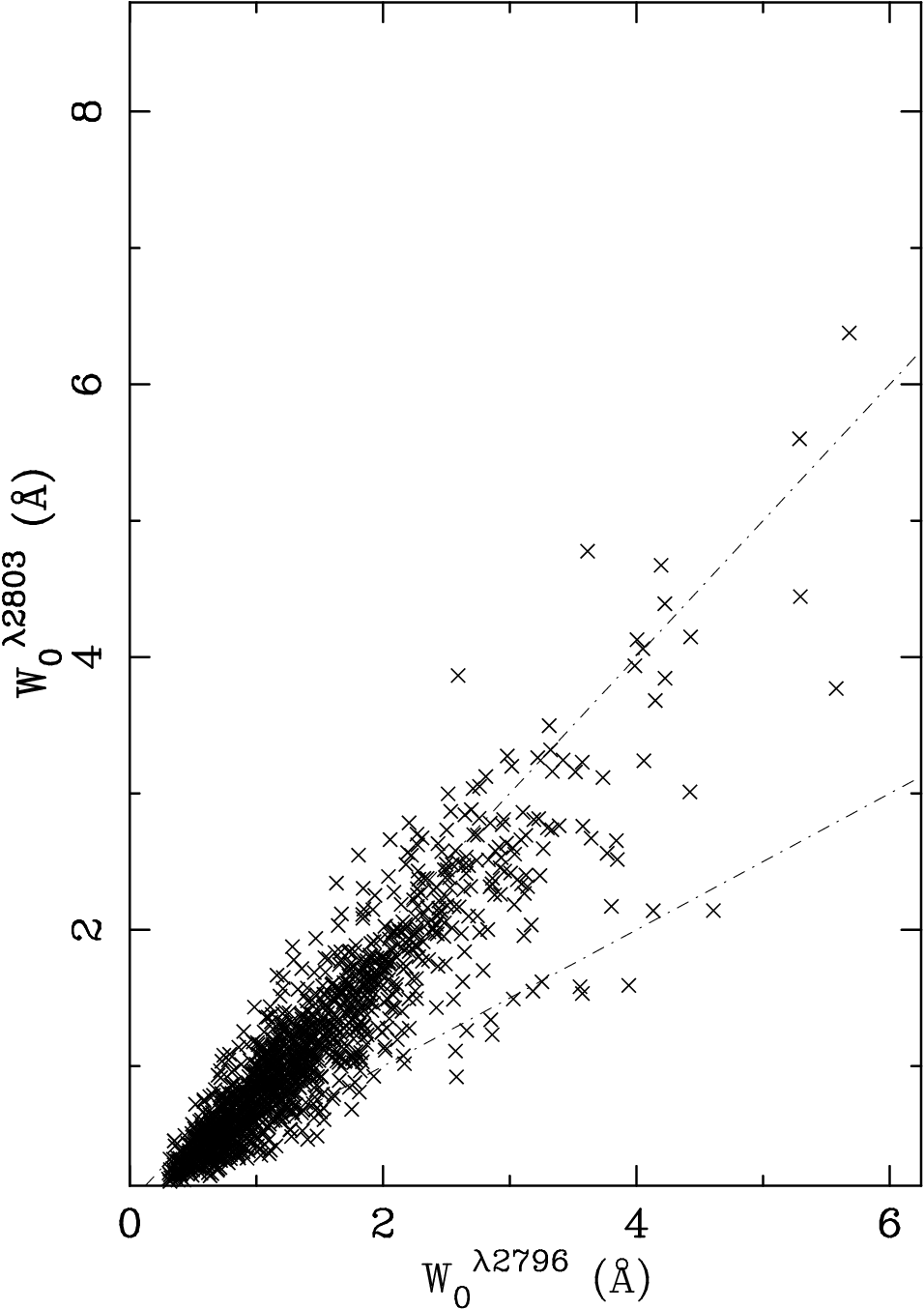}
\caption{Left: the distribution of
$W_0^{\lambda2796}/W_0^{\lambda2803}$ doublet ratio for  \ion{Mg}{2}
systems with $W_0^{\lambda2796} \ge 0.3$\,\AA.   The dashed lines mark
the limits of 1.0 for completely saturated systems and 2.0 for
completely unsaturated systems.  Values above and below these limits
are due to noise.  Our sample is dominated by saturated systems.
Right:  $W_0^{\lambda2803}$ versus $W_0^{\lambda2796}$.}
\label{fig_dr}
\end{figure}

\begin{figure}
\plotone{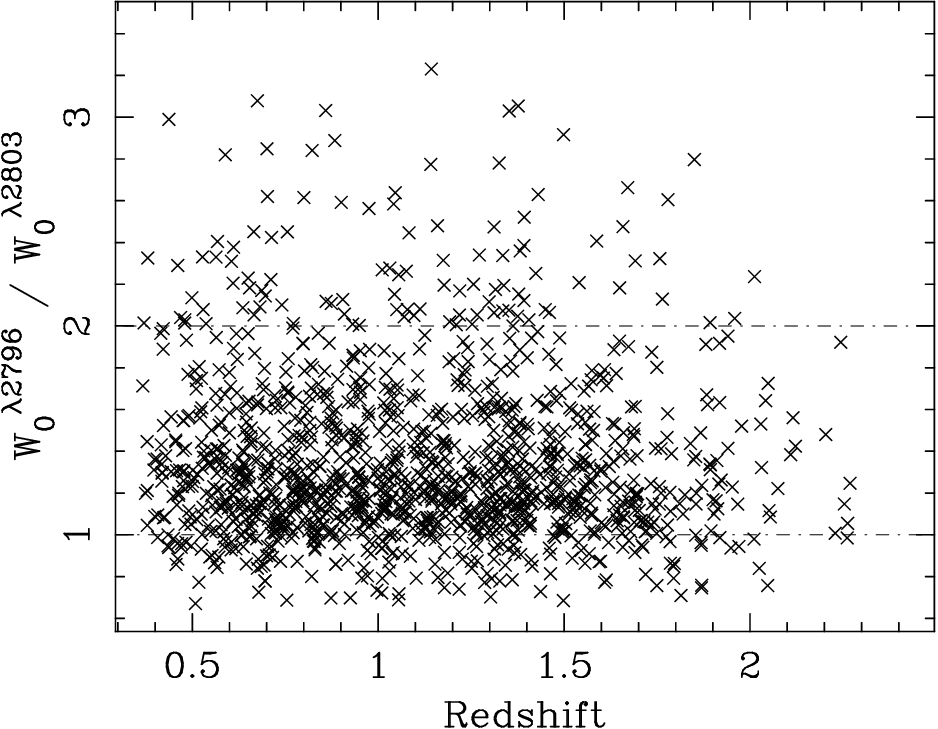}
\caption{$W_0^{\lambda2796}/W_0^{\lambda2803}$ doublet ratio as a
function of redshift for \ion{Mg}{2} systems with  $W_0^{\lambda2796}
\ge 0.3$\,\AA.  There is no detected redshift evolution in the doublet
ratio.}
\label{fig_zdr}
\end{figure}

\ion{Mg}{2} doublet ratios span the range from $DR=2.0$ for completely
unsaturated systems to  $DR=1.0$ for completely saturated systems.
Figure \ref{fig_dr} shows the $DR$ distribution for our sample.  The
doublets are, for the most part, saturated.  Ninety percent of
doublets with a measured $DR$ are within $3\sigma$ of $DR = 2.0$, and
92\% have $DR - 1.0 > \sigma_{DR}$.  Figure \ref{fig_zdr} shows $DR$
as a function of redshift.   There is no detectable evolution in the
$DR$ distribution.  In fact, the $DR$ distribution is remarkably
consistent over the three redshift ranges of Figures \ref{fig_rewd_3b}
and \ref{fig_dwdz}.

\subsection{\ion{Mg}{2} Velocity Dispersions}
\begin{figure}
\plotone{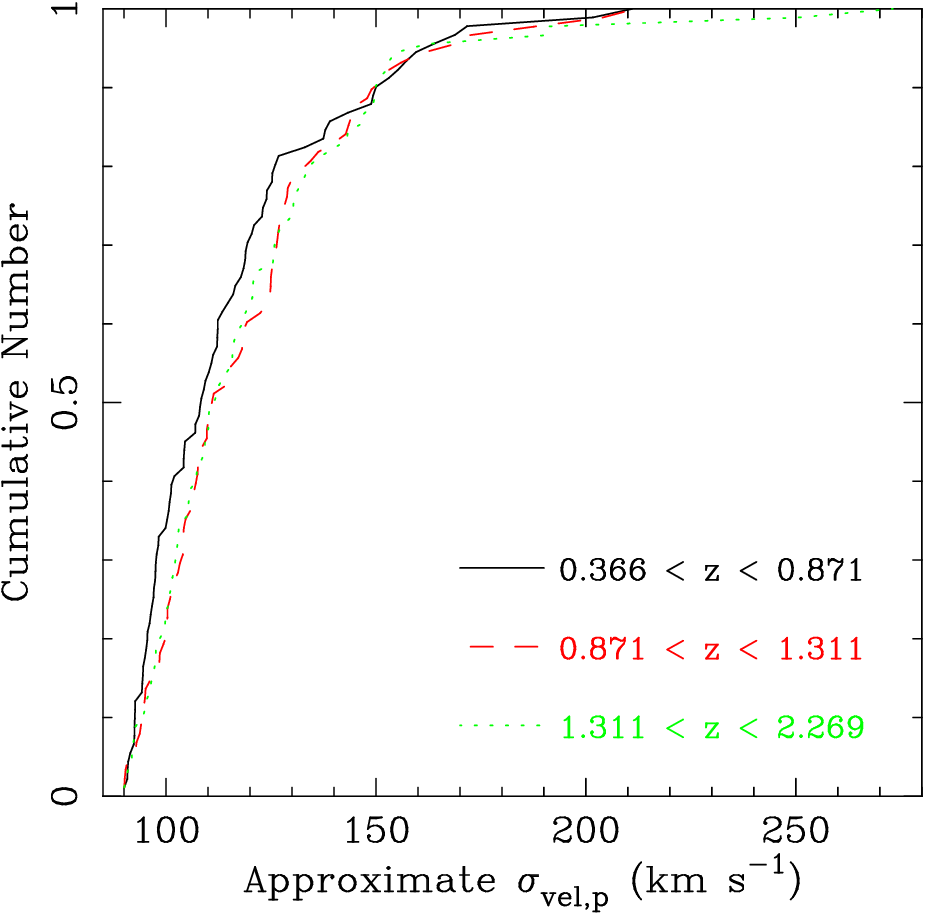}
\caption{Cumulative distribution of approximate velocity dispersions
for  $\sigma_{vel,p} > 90$ km s$^{-1}$.  The lower and middle redshift
bins differ in K-S probability at the $\approx1\sigma$ level.}
\label{fig_vel}
\end{figure}

For absorption lines that are at least partially saturated, such as
those that dominate our sample, $W_0^{\lambda2796}$ is primarily a
measure of the number of kinematic subcomponents along the line of
sight (Petitjean \& Bergeron 1990) and, to a lesser extent, projected
velocity dispersion, $\sigma_{vel,p}$ (see, for example, Churchill et
al. 2000, hereafter C2000).  We can directly extract information about
$\sigma_{vel,p}$ for individual systems by deconvolving the fitted
profile from the line spread function.  However, as most of the
$\lambda2796$  lines are only mildly resolved, this approach is not
sensitive for all but the most resolved systems.  Nonetheless, we
investigated the evolution of the $\sigma_{vel,p}$ distribution over
the three redshift  ranges of Figures \ref{fig_rewd_3b} and
\ref{fig_dwdz}.  In all three redshift ranges, $\simeq 80\%$ of the
systems  have measured values of $\sigma_{vel,p} < 90$ km s$^{-1}$.
Comparing just systems with $\sigma_{vel,p} > 90$ km s$^{-1}$, we find
that $\sigma_{vel,p}$ tends to be smaller in the lowest redshift bin
(see Figure \ref{fig_vel}).  However, the middle- and lower-redshift
bin velocity distributions only differ at the $1\sigma$ level.

\subsection{\ion{Fe}{2} $\lambda2600$ and \ion{Mg}{1} $\lambda2852$}
\begin{figure}
\plottwo{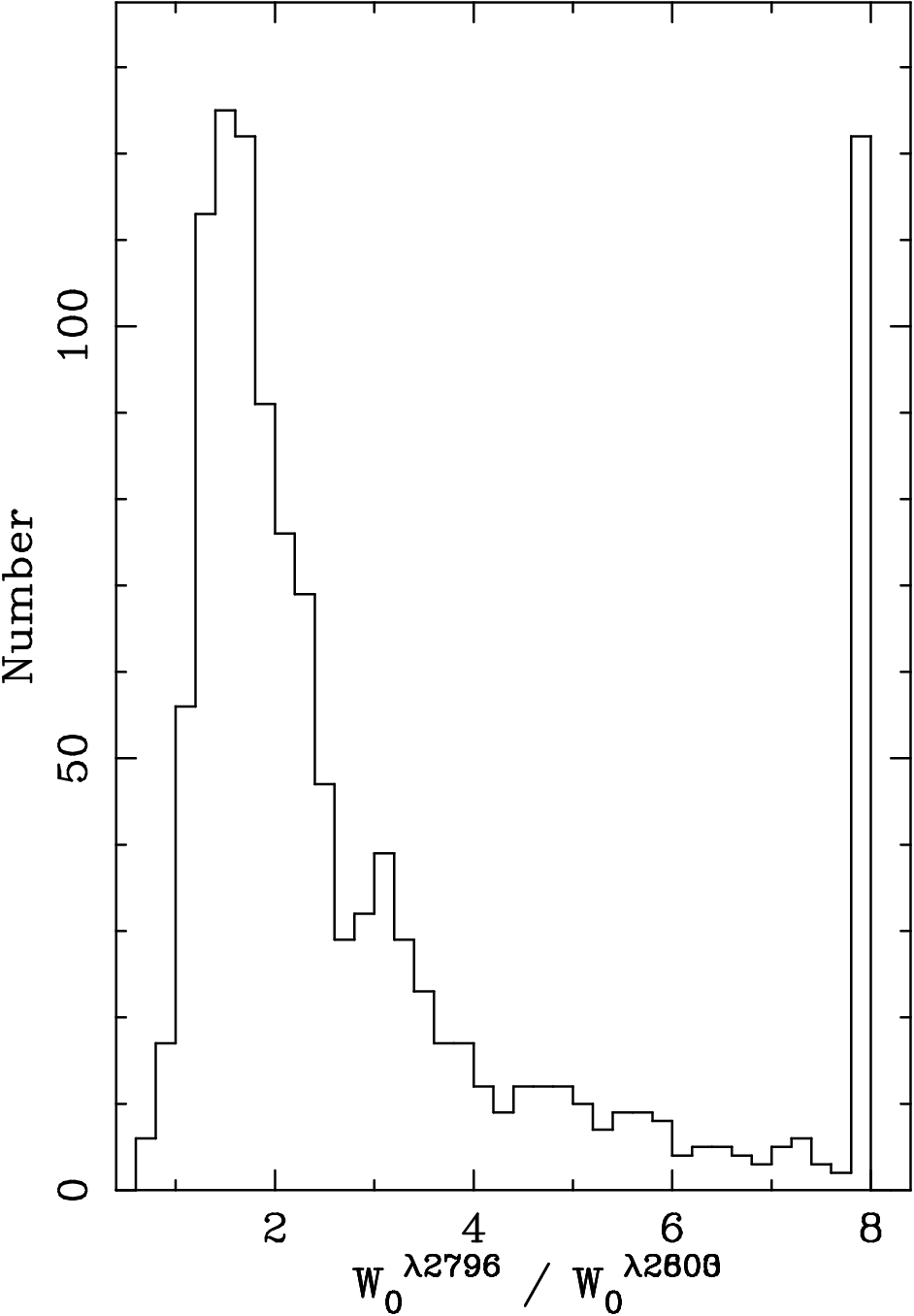}{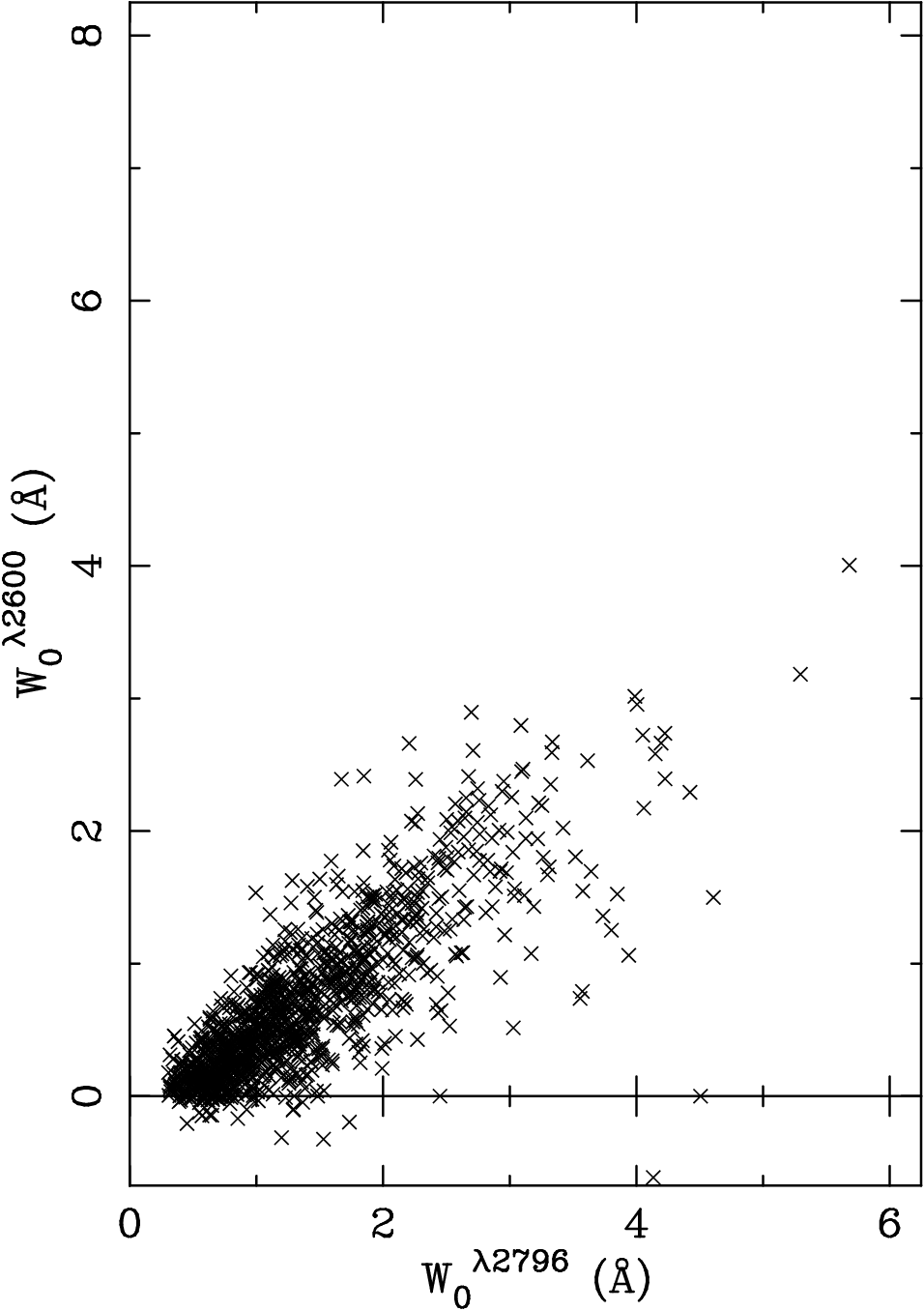}
\caption{\ion{Mg}{2} -- \ion{Fe}{2} ratio.   Left: The distribution of
\ion{Mg}{2} $\lambda2796$ -- \ion{Fe}{2} $\lambda2600$  line ratios.
The largest bin also contains all systems with ratios greater than
8.0, which includes non-detections of \ion{Fe}{2} $\lambda2600$.  The
mean is $\left<W_0^{\lambda2796}/W_0^{\lambda2600}\right> = 1.42$ and
ratios above $\approx 4$ are dominated by values with significance
less than $3\sigma$.  Right:  $W_0^{\lambda2600}$ versus
$W_0^{\lambda2796}$.  Negative  $W_0^{\lambda2600}$ values are
non-detections with scatter below zero.}
\label{fig_drfe}
\end{figure}

\begin{figure}
\plottwo{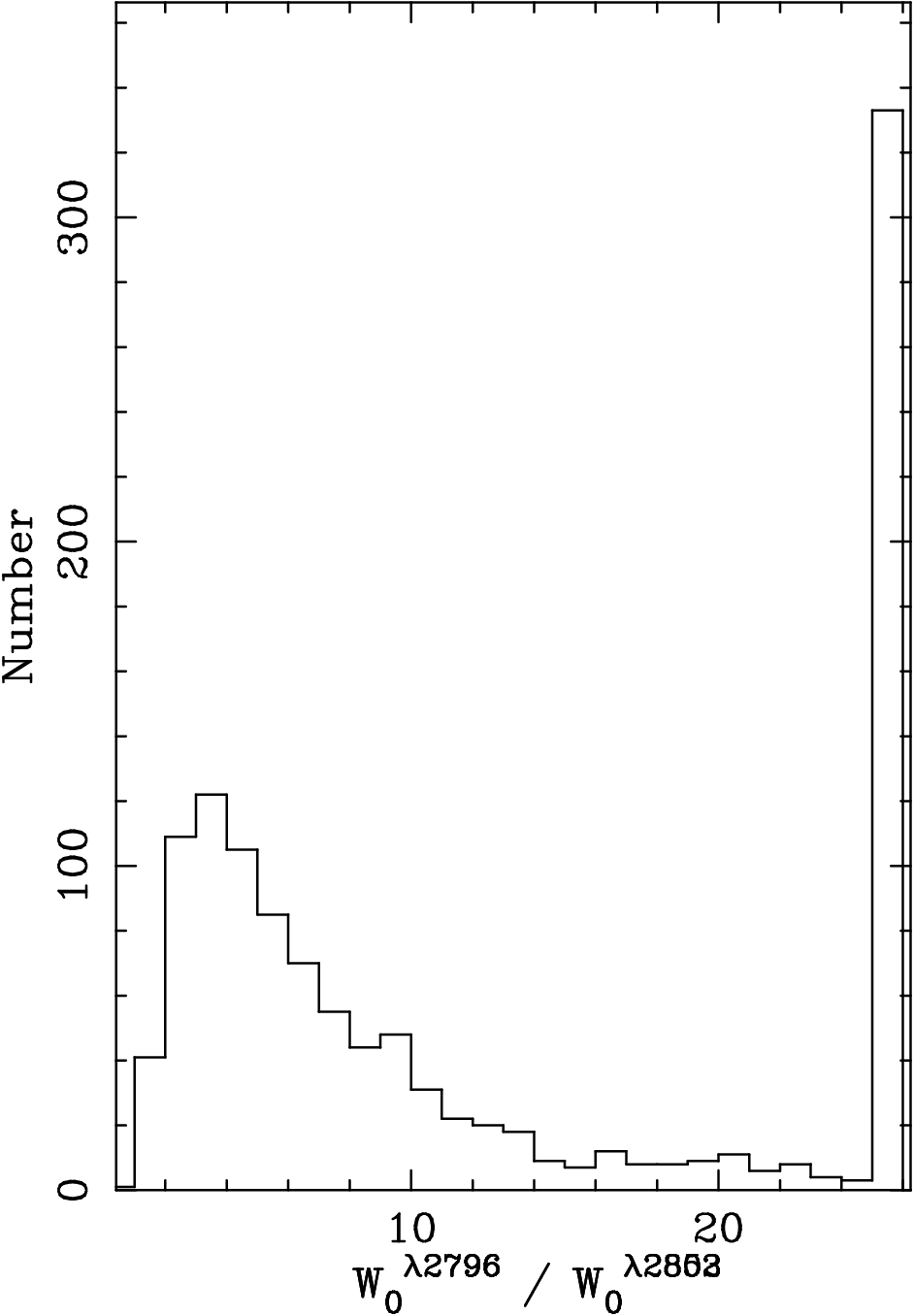}{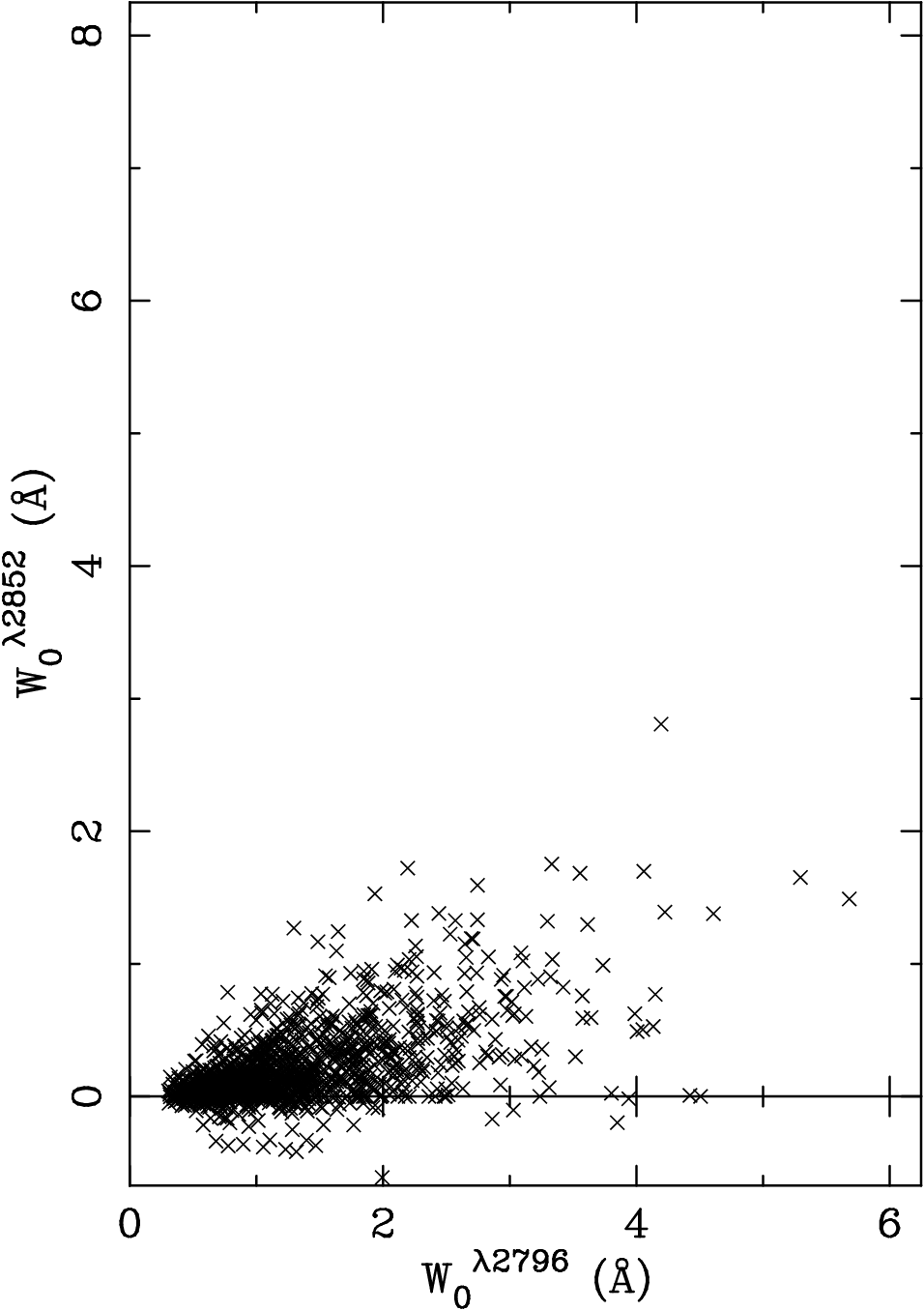}
\caption{\ion{Mg}{2} -- \ion{Mg}{1} ratio.   Left: The distribution of
\ion{Mg}{2} $\lambda2796$ -- \ion{Mg}{1} $\lambda2852$  line ratios.
The largest bin also contains all systems with ratios greater than 25,
which includes non-detections of \ion{Mg}{1} $\lambda2852$.  The mean
is $\left<W_0^{\lambda2796}/W_0^{\lambda2852}\right> = 4.14$ and
ratios above $\approx 8$ are dominated by values with significance
less than $3\sigma$.  Right:  $W_0^{\lambda2852}$ versus
W$_0^{\lambda2796}$.  Negative  $W_0^{\lambda2852}$ values are
non-detections with scatter below zero.}
\label{fig_drmg1}
\end{figure}

Figure \ref{fig_drfe} shows the distribution of \ion{Fe}{2}
$W_0^{\lambda2600}$ for our sample.  The error-weighted mean is
$\left<W_0^{\lambda2796}/W_0^{\lambda2600}\right> = 1.42$  and the
distribution has a smooth tail extending out to large values, though
ratios above $\approx 4$ are dominated  by values with significance
less than $3\sigma$.  Figure \ref{fig_drmg1} shows the distribution of
\ion{Mg}{1} $W_0^{\lambda2852}$ for our sample.  The error-weighted
mean is $\left<W_0^{\lambda2796}/W_0^{\lambda2852}\right> = 4.14$  and
the distribution has a smooth tail extending out to large values,
though ratios above $\approx 8$ are dominated  by values with
significance less than $3\sigma$.

\section{Discussion}
The determination of $\partial^2 N/\partial z\, \partial
W_0^{\lambda2796}$   provides insight into the nature of the systems
comprising the population of \ion{Mg}{2} absorbers.  Specifically, the
inability of a single functional form to describe $\partial N/\partial
W_0^{\lambda2796}$ suggests that multiple physical populations
contribute to the absorption.  Additionally, $\partial N/\partial
W_0^{\lambda2796}$ can be used to infer average absorption cross
sections for a given $W_0^{\lambda2796}$ regime.  The
$W_0^{\lambda2796}$-dependent evolution of $\partial N/\partial z$
holds further clues related to the nature of \ion{Mg}{2} absorbers.
These issues are discussed below.  We also note that a significant
fraction ($\approx$ 36\%) of \ion{Mg}{2} systems with
$W_0^{\lambda2796}\ge 0.5$\,\AA\ and $W_0^{\lambda2600}\ge 0.5$\,\AA\
are DLAs.  The implications of the evolution in the \ion{Mg}{2}
$\partial N/\partial z$ to $\Omega_{DLA}$ will be discussed in  Rao,
Turnshek, \& Nestor (2004).

\subsection{The Nature of the Absorbers}
\label{sec_nature}
The distribution of $W_0^{\lambda2796}$ for absorbers with
$W_0^{\lambda2796}\ge0.3$\,\AA\ is fit very well by a single
exponential.  However, extrapolating this exponential to smaller
values of  $W_0^{\lambda2796}$ under-predicts the incidence of lines
(Figure \ref{fig_rewd}), motivating a two-component description of the
form of Equation \ref{eqn2}.  This is the first clear indication of
such a transition and raises the question of whether the ensemble of
clouds comprising \ion{Mg}{2} absorbers  are the result of two
physically distinct populations.

In a series of papers using HIRES/Keck data  (Charlton \& Churchill
1998; C2000; Churchill \& Vogt 2001; Churchill, Vogt, \& Charlton
2003), Churchill and  collaborators have investigated the kinematic
structure of \ion{Mg}{2} absorbers at $0.4 < z \lesssim 1.2$.  For
intermediate/strong absorbers ($W_0^{\lambda2796}\ge0.3$\,\AA), they
show that the absorption systems are composed of multiple kinematic
subsystems, usually containing a dominant subsystem (ensemble of
clouds) with a velocity width $\approx 10-15$\,km\,s$^{-1}$ and a
corresponding \ion{H}{1} column density which is optically thick at
the Lyman limit  (i.e., $N_{H\,I} \gtrsim 3 \times 10^{17}$ atoms
cm$^{-2}$).  Additionally, a number of weaker, often unresolved (at
$\approx$6 km s$^{-1}$ resolution) subsystems (clouds) with a large
spread of velocity separations (up to $\approx 400$\,km\,s$^{-1}$)
from the systemic velocity of the absorber are usually present (see
Churchill \& Vogt 2001, Figure 7.)  These weaker subsystems typically
contribute only 10\%-20\% of the rest equivalent width.  The strongest
\ion{Mg}{2} absorbers ($W_0^{\lambda2796}\gtrsim2$\,\AA) often contain
more than one of the dominant subsystems and have equivalent widths
dominated by saturated features.

CRCV99 presented a HIRES/Keck study of weak
($W_0^{\lambda2796}<0.3$\,\AA) \ion{Mg}{2} absorption systems.  The
properties of the kinematic subsystems comprising weak \ion{Mg}{2}
absorbers are similar to the weak subsystems of intermediate/strong
\ion{Mg}{2} absorbers.  The individual clouds have
$W_0^{\lambda2796}\lesssim0.15$\,\AA\ and sub-Lyman limit \ion{H}{1}
columns.

Although most of the \ion{Mg}{2} $\lambda2796$ lines in our survey are
at least partially saturated, the {\it degree} of saturation is
correlated with $W_0^{\lambda2796}$.  For example, the DR values for
systems with 0.3\,\AA\ $\lesssim W_0^{\lambda2796} \lesssim 1$\,\AA\
indicate a mix in degree of saturation.  For lines in this regime,
$\partial N/\partial W_0^{\lambda2796}$ is sensitive to the
distribution  of \ion{Mg}{2} column densities, $\partial N/\partial
N_{Mg\,II}$.   In fact, Churchill \& Vogt (2001) and  Churchill, Vogt,
\& Charlton (2003) find almost identical  power law slopes for
$\partial N/\partial W_0^{\lambda2796}$ and  $\partial N/\partial
N_{Mg\,II}$ in their 0.3\,\AA\ $\le W_0^{\lambda2796} \le 1.5$\,\AA\
sample.  Systems with $W_0^{\lambda2796}\gtrsim2$\,\AA, however, are
almost always highly saturated.  Also, a {\it general} distinction
between weak ($W_0^{\lambda2796} < 0.3$\,\AA) and intermediate/strong
($W_0^{\lambda2796} \ge 0.3$\,\AA) \ion{Mg}{2} absorbers is the
absence or presence of a ``dominant'' subsystem, though the
differences are neither discrete nor without exception.  Additionally,
Rigby, Charlton, \& Churchill (2002) note a substantial excess of
\ion{Mg}{2} absorbers that are comprised of a single kinematic
component, which are almost exclusively weak absorbers.  Therefore,
given the form of $\partial N/\partial W_0^{\lambda2796}$, it is
appropriate to consider a picture in which the {\it sum} of the weaker
kinematic subsystems (clouds) in an absorber are described by
$N^*_{wk}$ and $W^*_{wk}$, and the stronger (dominant ensembles)
subsystems by $N^*_{str}$ and $W^*_{str}$.  If this
multi-population explanation is indeed correct, the nature of these
populations needs to be explained.  First, we consider the weak
subsystems.

It has been demonstrated that absorption line systems can be
associated with galaxies out to large galactocentric distances
($\approx 100 h^{-1}$ kpc  for \ion{C}{4} and $\approx 200 h^{-1}$ kpc
for  Ly$\alpha$ absorbers, for example; Chen, Lanzetta, \& Webb 2001;
Chen et al. 2001).  Though there is evidence for Ly$\alpha$ absorbers
in voids (Penton, Stocke, \& Shull 2002), these appear to be limited
to the weakest systems.  Also, Rigby, Charlton, \& Churchill (2002)
conclude that a significant fraction of relatively strong
($10^{15.8}$ cm$^{-2} \lesssim$ $N_{H\,I} \lesssim 10^{16.8}$
cm$^{-2}$) Ly$\alpha$ forest lines are associated with single-cloud weak \ion{Mg}{2}
absorbers, which are also \ion{C}{4} absorbers.

\begin{figure}
\plotone{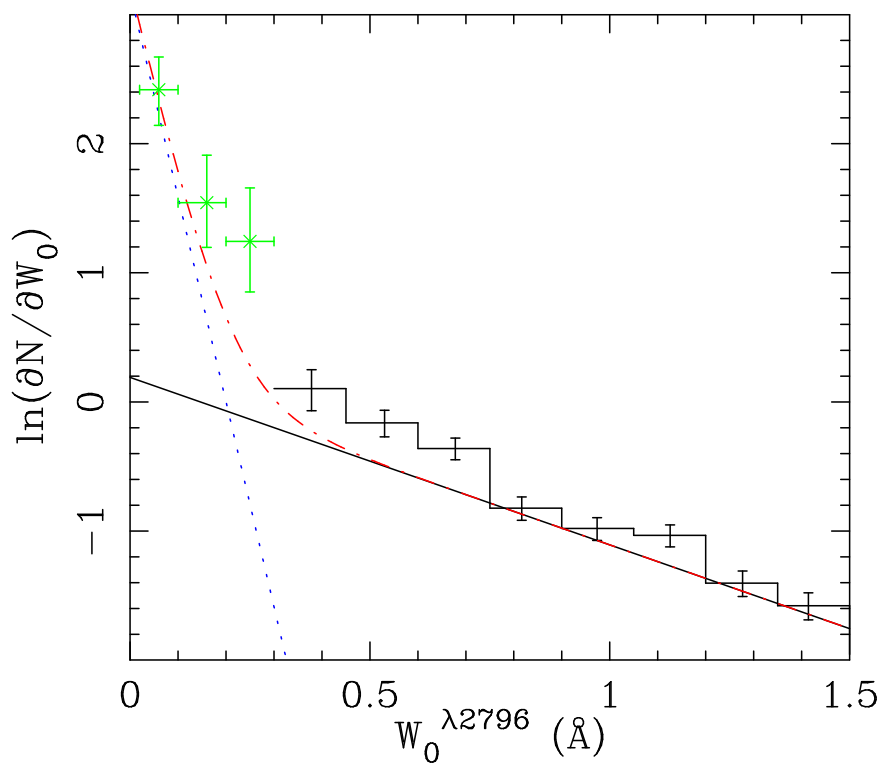}
\caption{Predicted $\partial N/\partial W_0^{\lambda2796}$ for
Ly$\alpha$ forest/single-cloud \ion{Mg}{2} .  The solid line is the
maximum likelihood fit to our data with $W_0^{\lambda2796}
\ge0.5$\,\AA.  The dotted line is the approximation for single-cloud
\ion{Mg}{2} $\partial N/\partial W_0^{\lambda2796}$ using data from
CRCV99 and a normalization from Rigby, Charlton, \& Churchill (2002).
The dot-dash curve is the sum of the two distributions.  The points
are from CRCV99.}
\label{fig_lyaf}
\end{figure}

Therefore, we consider whether single-cloud/Ly$\alpha$ forest
absorbers at relatively large galactocentric distances can contribute
significantly to our ``weak'' component of $\partial N/\partial
W_0^{\lambda2796}$.  CRCV99 claim to be 91\% complete down to
$W_0^{\lambda2796} = 0.03$\,\AA.  Thus, using the measurements of
single-cloud systems from CRCV99 with $W_0^{\lambda2796} \ge
0.03$\,\AA, we find $W^*_{sc} = 0.063 \pm 0.015$\,\AA.  Rigby,
Charlton, \& Churchill (2002) give a $\partial N/\partial z$ value for
the single-cloud systems, which can be used to normalize $\partial
N_{sc}/\partial W_0^{\lambda2796}$.   While our comparison is
approximate in that the redshift range covered by the C2000 sample,
$0.4 < z < 1.4$, is different than that of our sample, the result
shown in  Figure \ref{fig_lyaf} demonstrates that the
single-cloud/Ly$\alpha$ forest systems should contribute
significantly to the upturn in $\partial N/\partial
W_0^{\lambda2796}$, especially for the weakest systems.

Although the \ion{Mg}{2} absorber galaxy population has been shown to
span a range of galaxy colors and types, Steidel, Dickinson, \&
Persson (1994) describe the ``average'' galaxy associated with an
intermediate/strong \ion{Mg}{2} absorber galaxy as one which is
consistent with a typical $0.7L^*$ Sb galaxy.  Furthermore, Steidel at
al. (2002, also see Ellison, Mall\'{e}n-Ornelas, \& Sawicki 2003)
compare the galaxy rotation curves  to the absorption kinematics for a
sample of high inclination spiral \ion{Mg}{2}  absorption galaxies.
They find extended-disk rotation dominant for the absorption
kinematics, though a simple disk is unable to explain the range of
velocities, consistent with a disk plus halo-cloud picture.  However,
there are several counter evidences to a rotating disk description for
the dominant subsystems.  The most important is that imaging often
fails to reveal a disk galaxy in proximity to the absorption line of
sight.  DLA galaxies, which are a subset of strong \ion{Mg}{2}
absorbing galaxies, are not dominated by classic spirals.  Of the 14
identified DLA galaxies summarized in  Rao et al. (2003), only six are
spirals.  Furthermore, systems that have been studied  without the
discovery of the absorbing galaxy despite deep imaging rule out at
least bright spirals (although disky-LSBs may still contribute.)
Bright galaxies near QSO sightlines usually show \ion{Mg}{2}
absorption, but not {\it all} strong \ion{Mg}{2} absorbers have a
nearby spiral or bright galaxy.  Thus, though it is clear that
rotating disks do contribute to the population comprising the dominant
kinematic subsystem(s) of $W_0^{\lambda2796} \ge 0.3$\,\AA\ absorbers,
they can only account for a (perhaps small) fraction of the total
population.

The detailed nature of the remaining contribution is yet unclear, but
this dichotomy supports the idea of multiple populations.   Bond et
al. (2001), for example,  show that the kinematics of many strong
(defined as $W_0^{\lambda2796} \ge 1.8$\,\AA\ in their sample)
\ion{Mg}{2} absorbers have kinematic structure that is highly
suggestive of superwinds/superbubbles.  They also show that DLAs
exhibit low-ion kinematic profiles distinct from the superwind-like
absorbers, which tend to be sub-DLA.  Furthermore, imaging studies (e.g., Le
Brun et al. 1997; Rao et al. 2003) have found  LSB, dwarf, and
groups/interacting systems, in addition to some spirals,  responsible
for strong \ion{Mg}{2} absorption systems (though the samples are
typically selected to be DLA sightlines).   Sightlines through groups
and interacting systems are likely to sample large velocity spreads.
Apparently, their contribution to the total cross section of strong
\ion{Mg}{2} systems is non-negligible.

Thus, though a two-component description is sufficient to describe
$\partial N/\partial W_0^{\lambda2796}$  within the current data, it
is likely that more than two physically distinct populations
contribute significantly to the phases comprising the absorbing
clouds.  By observing a large number of galaxy types that sample a
range of  $W_0^{\lambda2796}$ and impact parameter, it would be
possible to  ascertain what populations contribute most directly to
the absorber population.

\subsection{The Absorber Cross-Sections}
The relative mean absorber cross-section, $\sigma(z)$, for ranges of
$W_0^{\lambda2796}$ can be determined directly from $\partial
N/\partial W_0^{\lambda2796}$.  Additionally,  knowledge of  $\partial
N/\partial z$ and the galaxy luminosity function (LF) allows the
approximation of the normalization of the mean absorber cross-section,
under the assumption that the absorbers arise in galaxies,  since
$\partial N/\partial z$ is the product of $\sigma(z)$ and the number
density of absorbers, $n(z)$.  However, the process is inherently
uncertain since: (a) the redshift dependence of the galaxy LF is not
well determined; (b) although a scaling relation
$\sigma(L)\propto(L/L^*)^\beta$ has traditionally been assumed,  the
determination of $\beta$ is uncertain and likely not universal for all
morphological types or redshifts, if it is even appropriate; and (c)
depending on the value of $\beta$ and the LF faint-end slope, the
minimum absorber galaxy luminosity (which also is likely to be
dependent on morphology and redshift) can have large effects in the
statistically-determined cross sections.

Nonetheless, an approximate normalization is of physical interest.
Thus, we define $R_{eff}$ to be the effective projected radius for
absorption such that $\sigma(L) \equiv \pi\,R_{eff}^2(L)$.   Note that
the galactocentric radius within which the absorption may occur (i.e.,
the range of impact parameter) may be very different than $R_{eff}$.
Steidel, Dickinson, \& Persson (1994) find absorbing galaxies as faint
as $L \sim 0.05\,L^*$, and a scaling law for the gaseous cross-section
$R_{eff}(L)/R_{eff}^* = (L/L^*)^{0.2}$ for both optical (B-band) and
near-IR (K-band) luminosity.  They claim the relation is much tighter
for the near-IR.  Thus, we use the redshift parameterization of the
K-band LF from the  MUNICS data set (Drory et al. 2003) and $L_{min} =
0.05\,L^*$, though we also test $L_{min} = 0.25\,L^*$ and $L_{min} =
0.001\,L^*$ to investigate the effects of using high/low values of
$L_{min}$.  We adopt the Dickinson \& Steidel (1996) revised scaling
law for \ion{Mg}{2} systems  with $W_0^{\lambda2796} \ge 0.3$\,\AA,
namely $R_{eff}(L_K)/R_{eff}^* = (L_K/L_K^*)^{0.15}$.

\begin{figure}
\plotone{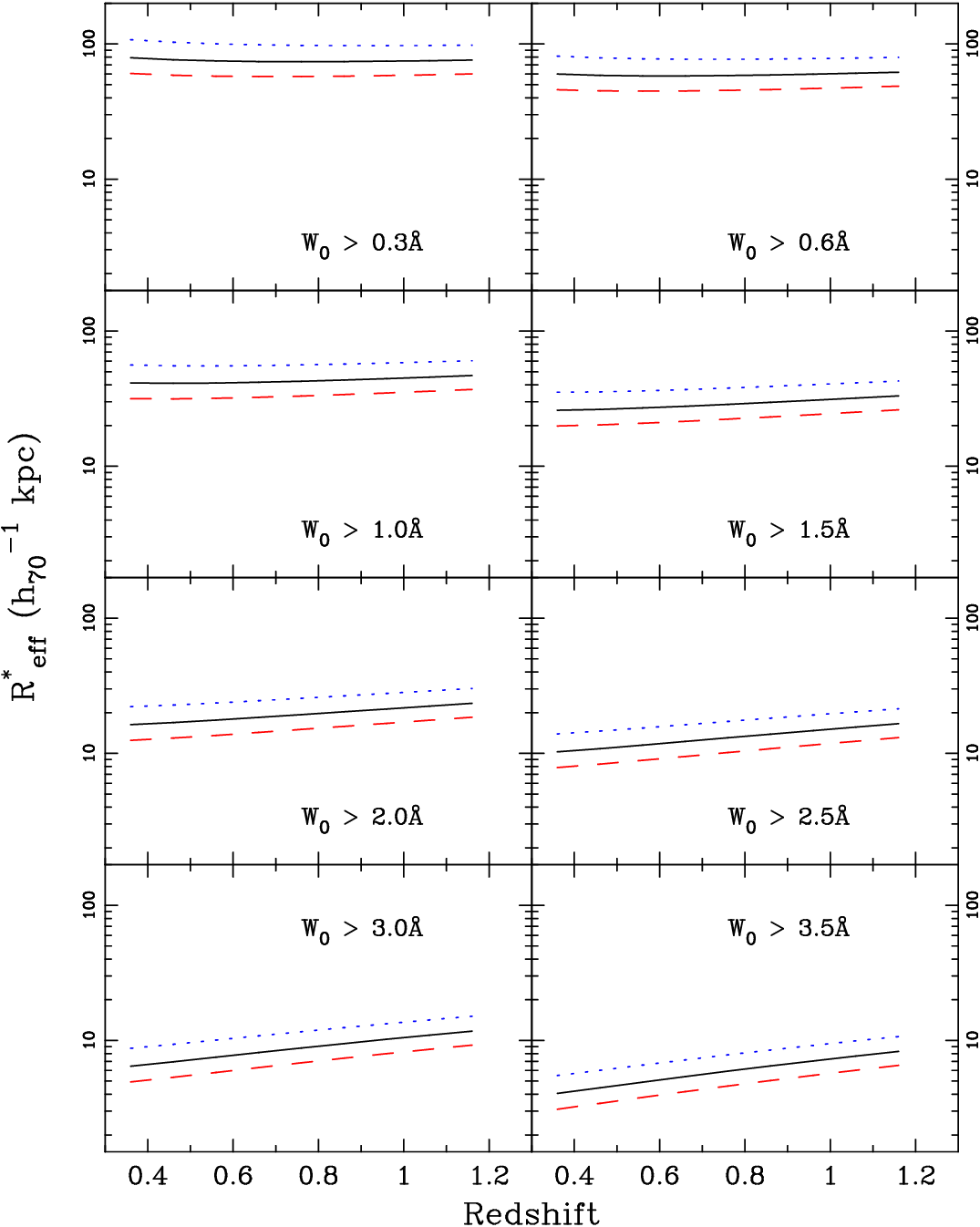}
\caption{Cross-section estimates $\sigma(L^*) = \pi\,R_{eff}^{*\,2}$.
The solid lines represent the $L_{min}=0.05\,L^*$ result.  The
dashed-lines are for a small value of $L_{min}=0.001\,L^*$ and the
dotted-lines a large value $L_{min}=0.25\,L^*$.  We use the redshift
parameterization of the K-band LF from the  MUNICS data set, a scaling
relation $R_{eff}/R_{eff}^* = (L/L^*)^{0.15}$, and $\partial
N/\partial z$ from our parameterization (\S \ref{sec_wzd}).}
\label{fig_cs}
\end{figure}

The results for the different  $W_0^{min}$ values from \S3.2 are shown
in Figure \ref{fig_cs}.  Despite the limitations of this approach,
approximate values for $R^*_{eff}$ are apparent: $\approx
60-100\,h_{70}^{-1}$\,kpc for  $W_0^{\lambda2796} \ge 0.3$\,\AA,
$\approx 30-60\,h_{70}^{-1}$\,kpc for  $W_0^{\lambda2796} \ge
1.0$\,\AA, $\approx15-30\,h_{70}^{-1}$\,kpc  for $W_0^{\lambda2796}
\ge 2.0$\,\AA\ and $\lesssim 10\,h_{70}^{-1}$\,kpc for
$W_0^{\lambda2796} \ge 3.5$\,\AA.   Guillemin \& Bergeron (1997) find
$R^*_{eff} = 65\,h_{70}^{-1}$ kpc using the upper envelope of the
distribution of observed impact parameters for a sample with 0.3\,\AA\
$\le W_0^{\lambda2796} \le 4.7$\,\AA\  and
$\left<W_0^{\lambda2796}\right> = 1.3$\,\AA.  CRCV99 find $R^*_{eff} =
110\,h_{70}^{-1}$ kpc for their $W_0^{\lambda2796} \ge 0.02$\,\AA\
sample, but use a LF normalization that is $\sim 3.5$ times our
adopted values.  Chen, Lanzetta, \& Webb (2001) find $R^*_{eff} =
137\,h_{70}^{-1}$ kpc for \ion{C}{4} absorbers, which may sample
larger  galactocentric radii.  The $R^*_{eff}$ values are averages; it
is expected that there are  large galaxy to galaxy variations.  Direct
comparison of $R^*_{eff}$ to the distribution of  impact parameters
would require knowledge of the gas geometry and the  (likely
$W_0^{\lambda2796}$-dependent) covering factor.  For example, small
regions that allow large $W_0^{\lambda2796}$ absorption may be found
at galactocentric radii $r \gg R_{eff}$ if the covering factor for
such absorption is $f \ll 1$.

We also note that for the single-cloud \ion{Mg}{2} systems (see
\S\ref{sec_nature}), Rigby, Charlton, \& Churchill (2002) derive
$\partial N/\partial z=1.1\pm0.06$.  They claim that these systems are
25\%-100\% of the Ly$\alpha$ forest with $10^{15.8}$ cm$^{-2} <$
$N$(\ion{H}{1}) $< 10^{16.8}$ cm$^{-2}$.  Under the assumption that they
can be associated with galaxies, their incidence corresponds to
$R^*_{eff} \approx 80 - 120\,h_{70}^{-1}$ kpc in the above approximation.
This can be compared to the claim of Chen et al. (2001) that all
$N$(\ion{H}{1}) $\gtrsim 10^{14}$ cm$^{-2}$ Ly$\alpha$ forest lines arise
within the characteristic radius $R^*\approx250\,h_{70}^{-1}$ kpc of
luminous galaxies.

\subsection{Nature of the Evolution}
\subsubsection{$W_0^{\lambda2796} < 0.3$\,\AA\ Systems}
The Ly$\alpha$ forest $\partial N/\partial z$ is known to decrease
strongly over the redshift interval $4 \gtrsim z \gtrsim 1.6$, but
little evolution is present at lower redshifts (see Weymann 1998, and
references therein).  In \S\ref{sec_nature}, we speculated that a
portion of the enriched Ly$\alpha$ forest contributes to the weak
subsystems found in \ion{Mg}{2} absorbers.  If this is indeed the
case, then this evolution may manifest itself in $\partial N/\partial
z$ for the weakest ($W_0^{\lambda2796} < 0.3$\,\AA) \ion{Mg}{2}
systems at high $z$.

\subsubsection{The Lack of Measured Evolution in Intermediate/Strong (0.3\,\AA\ $\le W_0^{\lambda2796} \lesssim 2$\,\AA) Systems}
\label{sec_stab}
Our sample covers the redshift interval $z=2.267$ to $z=0.366$ which
corresponds to 6.6 Gyrs, or about half the age of the universe.  As
long assumed and recently demonstrated with the Hubble deep fields
(see Ferguson, Dickinson, \& Williams 2000 for a review), galaxies
were quite different at $z=2.3$ than at $z=0.4$.  It is therefore
noteworthy that, even with the large absorber sample presented here,
relatively little evolution is detected in the bulk of \ion{Mg}{2}
absorption systems.  For example, it is believed that the global star
formation rate peaked near $z=1$ and has declined by a factor of
$\approx2$ by $z=0.5$ and by a  factor of $\approx10$ by $z=0$
(Hopkins 2004 and references therein.)  Contrastingly, the total
cross-section for absorption  $n(z)\,\sigma(z)$ of  our $W_0^{min} =
0.6$\,\AA\ and 1.0\,\AA\ samples evolves $\lesssim 10\%$ from $z=1.9$
to $z=0.5$ (corresponding to an interval of 5 Gyrs).  The
corresponding $R_{eff}$ values do, however, indicate that much of this
cross section is at large galactocentric radius, extending well beyond
stellar galactic radii.

Two immediate conclusions can be drawn from the lack of evolution in
the total cross sections.  First, a large majority of the structures
responsible for the  bulk of the absorption cross-section were in
place by $z \approx 2$.  Second, either the time scales governing the
dynamics of the structures is greater than several Gyrs, or some
process(es) regulate the production/destruction of the structures such
that a nearly steady state is reached.

Mo \& Miralda-Escud\'{e} (1996) explore two-phase models for gaseous
galactic halos in which ``cold phase'' clouds condense from the
extended ``hot phase'' ionized halo  at a cooling radius $r_c$, and
undergo infall at $v\simeq V_{cir}$.  Setting $r_c = R^*_{eff}$
(appropriate for a spherical geometry and a covering factor of unity),
and using $R^*_{eff} \simeq 100$ kpc, their model results give
$V_{cir} \simeq 250-400$\,km\,s$^{-1}$ over $2 \ge z \ge 1$, which is
consistent with the observed kinematic ranges over $1.2 \ge z \ge 0.4$
from Churchill \& Vogt (2001).  Using these values we obtain
$t_{infall} \approx 0.25 - 0.4$ Gyr.  Thus, it would appear that a
large degree of regulation is necessary, in this picture, to explain
the lack of  evolution in $n(z)\,\sigma(z)$.  Ionization from star
formation and the disruption from instabilities and evaporation of
smaller clouds should at least partially regulate the cooling.  The
result of these and/or other regulatory mechanisms must ensure that
the minimum time span  over which condensation occurs be equal to at
least several times $t_{infall}$.

Alternatively, lifetimes could indeed be much smaller than the time
interval here studied, but a range of formation epochs conspire to
maintain a roughly constant total cross section.  While this scenario
seems less likely than early formation epochs and long lifetimes
through regulation, it is of note that although individual low-mass
halos in $\Lambda$CDM simulations do themselves evolve, the low-mass end of the total halo mass
function shows little evolution in redshift (Reed et al. 2003).

The evolutionary situation for disks may be more complex.  Simulations
suggest that, to overcome ``angular momentum catastrophe'', local
disks were not formed until $z\le1$ (Mo, Mao, \& White 1998; Weil,
Eke, \& Efstathiou 1998) and have since undergone much evolution (Mao,
Mo, \& White, 1998).  Disks were likely   plentiful at higher $z$, but
smaller in size for a given $V_{cir}$, and largely dissipated or
destroyed in mergers.  Driver et al. (1998) find a deficit of $z>2$
spirals in the HDF suggesting that this marks the onset of their
formation.  Disks certainly contribute at least partially to the
absorber population, as suggested by Charlton \& Churchill (1998) and
confirmed for specific cases by Steidel et al. (2001).  However, even
if their contribution to  0.3\,\AA\ $\le W_0^{\lambda2796} \lesssim
2.0$\,\AA\ \ion{Mg}{2} absorption systems is significant, the lack of
evolution remains difficult to explain.

\subsubsection{Evolution of Ultra-Strong Systems ($W_0^{\lambda2796} \gtrsim2$\,\AA)}
We detect  evolution in $W_0^{\lambda2796}$ \ion{Mg}{2}
systems, in the sense that the strongest lines evolve away, with the
evolution being stronger for increasingly stronger lines and at
redshifts $z \lesssim 1$.  For example, we find that the total
cross-section for absorption of \ion{Mg}{2} systems with
$W_0^{\lambda2796} \ge 2.0$\,\AA\ ($\approx10\%$ of all systems with
$W_0^{\lambda2796} \ge 0.3$\,\AA) decreases by $45^{+14}_{-17}\%$ from
$z=1.8$ to 0.6.  The total cross-section for absorption of systems
with $W_0^{\lambda2796} \ge 3.5$\,\AA\ ($\approx1\%$ of all systems
with $W_0^{\lambda2796} \ge 0.3$\,\AA) decreases by $60^{+19}_{-34}\%$
from $z=1.9$ to 0.8.   Note that for the $W_0^{\lambda2796}$ values
considered in this work, $W_0^{\lambda2796}$ is correlated with the
number of kinematic subsystems (clouds) comprising the absorber and
thus velocity dispersion.

A speciously simple picture for the evolution of \ion{Mg}{2}
$\lambda2796$ lines considers that velocity dispersions scale with
halo mass, and halo masses grow with decreasing  redshift.  In this
picture, with the approximation that the cross-section for a given
dispersion scales with mass and the mass spectrum evolves according to
the formalism of Sheth and Tormen (Sheth \& Tormen 1999), the
cross-sections should {\it increase} with decreasing redshift by
approximately an order of magnitude per unit redshift for $2 > z > 0$.
This is clearly ruled out by the data.  This can be understood on
physical grounds if the low-ionization gas giving rise to \ion{Mg}{2}
absorption is in clouds bound in galaxy halos, since the individual
clouds are not expected to be virialized on  group scales.  A single
\ion{Mg}{2} system sampling a group dispersion would require  the
chance alignment of several virialized member galaxy halos along the
line of sight.

However, C2000 show kinematic profiles of four $z>1$ and one
$z\approx0$ system with $W_0^{\lambda2796} \ge 2$\,\AA, each of which
exhibit two strong ``dominant subsystem''  components, and suggest the
possibility that the strongest systems have a connection to galaxy
pairs.  For example, they find a 25\% chance that a random sightline
through the Galaxy would also intercept the LMC, and a 5\% chance that
it would intercept the SMC.  This possible connection can be
investigated by imaging $W_0^{\lambda2796} \ge 2$\,\AA\ systems: our
sample contains 77 such lines with $z<1$ and 20 with $z<0.6$.

Absorption $\partial N/\partial z$ is the product of the number
density of absorbers times the average individual cross-section for
absorption.  Most galactic halos are already in place by $z=1$, so
except for mergers of galaxies separated by $v > 500$\,km\,s$^{-1}$,
the evolution in $\partial N/\partial z$ is primarily an evolution of
the gas absorption cross-section in individual halos.  Evolution in
metallicity or metagalactic ionization cannot explain the
cross-section evolution,  since they would tend to increase the number
of individual  enriched, low ionization clouds along a line of sight
through a galaxy.  The $W_0^{\lambda2796}$-dependent decrease in cross
section must therefore be indicative of an evolution in the kinematic
properties of a subset of the galaxy population, from intermediate to
low redshift.  If interactions, galaxy pairs and
superwinds/superbubbles represent a significant fraction of the
ultra-strong \ion{Mg}{2} absorbers, then the decrease in the number of
interactions/pairs from high redshift (e.g., in the HDF) and the
decrease in the space density of superwinds (due to the
decrease in the global star formation rate at $z\lesssim1$) may in
part account for the evolution in  $\partial N/\partial z$.  Although
these issues and those discussed in \S\ref{sec_stab} likely hold clues
to the nature of this evolution, a precise description will require
knowledge of the physical nature of the clouds that comprise the
different ranges of $W_0^{\lambda2796}$ in \ion{Mg}{2} absorption
systems.

\section{Conclusions}
We have identified over 1,300 \ion{Mg}{2} absorption systems with
$W_0^{\lambda2796} \ge 0.3$\,\AA\ over the redshift range $0.367 \le z
\le 2.269$ in the SDSS EDR QSO spectra.  The size of the sample is
such that statistical errors are comparable to systematic effects and
biases.  We used simulations to improve our line-finding algorithm
until the systematics could no longer be isolated from the noise at a
level where we could further improve our parameterizations.   Using
the combined redshift and $W_0^{\lambda2796}$ data, we offered a new
redshift parameterization for the distribution of systems.  In
conclusion, we have shown:
\begin{enumerate}
\item{The rest equivalent width distribution for intervening
\ion{Mg}{2} $\lambda2796$ absorption lines detected in the SDSS EDR
QSO spectra with $W_0^{\lambda2796} \ge 0.3$\,\AA\  is very well
described by an exponential, with $N^*=1.187 \pm 0.052$ and $W^* =
0.702 \pm 0.017$\,\AA.  Power law parameterizations drastically
over-predict the number of strong lines, and our exponential
under-predicts previously reported values for the number density of
weaker lines.}
\item{When compared with the number density of  $W_0^{\lambda2796} <
  0.3$\,\AA\ lines from other studies, our results show that neither
  an exponential nor a power law accurately represents the full range
  of $W_0^{\lambda2796}$.  Simulations of our catalog support this
  finding.  We propose a combination of two exponential distributions,
  where the weak lines are described by  the parameters $N^*_{wk}
  \approx 1.7$ and $W^*_{wk} \approx 0.1$\,\AA\ and the moderate and
  strong lines by the parameters $N^*_{str} \approx 0.9$ and
  $W^*_{str} \approx 0.8$\,\AA.}
\item{The rest equivalent width distribution steepens with decreasing
redshift, with $W^*$ decreasing from $0.80\pm0.04$\,\AA\ at $z = 1.6$
to $0.59\pm0.02$\,\AA\ at $z=0.7$.}
\item{For lines with $W_0^{\lambda2796} \lesssim 2$\,\AA, there is no
significant evolution detected in $\partial N/\partial z$.}
\item{For lines with $W_0^{\lambda2796} \gtrsim 2$\,\AA, evolution is
detected in $\partial N/\partial z$, with a decrease from the
no-evolution prediction of $\approx 45\%$ from $z=1.8$ to 0.6.  The
evolution is stronger for stronger lines and redshifts $z \lesssim 1$.}
\item{The number density of \ion{Mg}{2} absorption lines with
$W_0^{\lambda2796} \ge 0.3$\,\AA\ is well parameterized by
$\frac{\partial ^2N}{\partial z\,\partial
W_0}=\frac{N^*(z)}{W^*(z)}\,e^{-\frac{W_0}{W^*(z)}}$, with
$N^*(z)=1.001\pm0.132\,(1+z)^{0.226\pm0.170}$ and
$W^*(z)=0.443\pm0.032\,(1+z)^{0.634\pm0.097}$\,\AA.}
\end{enumerate}
Several lines of evidence suggest that the clouds giving rise to
\ion{Mg}{2} absorption comprise multiple physically-distinct
populations.  The results from high resolution work and the apparent
transition in $\partial N/\partial W_0^{\lambda2796}$ are perhaps the
strongest evidences, but the apparent contribution of enriched
Ly$\alpha$ forest lines, the inability of simple physical models to
reproduce in full the kinematic data and the menagerie of galaxy
types, luminosities, environments and impact parameters that contribute to the
absorber galaxy population are also consistent with this picture.

The situation will be made even clearer with the analysis of the full
SDSS database.  With over an order of magnitude more data available,
there will be enough high signal-to-noise  ratio spectra for studies
to reach lines weaker than 0.3\,\AA\ providing coverage across the
$\partial N/\partial W_0^{\lambda2796}$  transition in a single
survey.  Although the large number of systems  will require large,
improved Monte Carlo simulations, the improved statistics will allow
for finer analysis of the strength-dependent redshift evolution.
Additionally, work currently  in progress will extend our analysis
down to redshifts $z\approx0.2$ and $W_0^{\lambda2796} \gtrsim0.1$
(Nestor et al., in preparation).

\acknowledgements The authors would like to thank Brice M\'{e}nard for
his insight concerning lensing and dust biases, Arie Barratt for
aid with the MINUIT software, Ravi Sheth for productive discussions
and insight concerning halo evolution, and the referee for valuable
comments and suggestions.  We also thank the members of the SDSS
collaboration who made the SDSS project a success and who made the EDR
QSO spectra  available.  We also acknowledge support from NASA-LTSA
and NSF.  Funding for the creation and distribution of the SDSS
Archive has been provided by the Alfred P. Sloan Foundation, the
Participating Institutions, the National Aeronautics and Space
Administration, the National Science Foundation, the U.S. Department
of Energy, the Japanese Monbukagakusho, and the Max Planck
Society. The SDSS Web site is http://www.sdss.org/.  The SDSS is
managed by the Astrophysical Research Consortium (ARC) for the
Participating Institutions. The Participating Institutions are The
University of Chicago, Fermilab, the Institute for Advanced Study, the
Japan Participation Group, The Johns Hopkins University, Los Alamos
National Laboratory, the Max-Planck-Institute for Astronomy (MPIA),
the Max-Planck-Institute for Astrophysics (MPA), New Mexico State
University, University of Pittsburgh, Princeton University, the United
States Naval Observatory, and the University of Washington.

\appendix
\section{$\frac{\partial ^2N}{\partial z\,\partial W_0}$ Parameterization}
In order to parameterize $\partial ^2N/\partial z\,\partial W_0$, we
write
\begin{equation}
\frac{\partial ^2N}{\partial z\,\partial W_0} =
\frac{N^*(z)}{W^*(z)}e^{-\frac{W_0}{W^*(z)}} =
\frac{\mathcal{N}^*}{\mathcal{W}^*}\,(1+z)^{\alpha-\beta}\,
e^{-\frac{W_0}{\mathcal{W}^*}(1+z)^{-\beta}},
\label{eqn}
\end{equation}
such that $N^*(z)=\mathcal{N}^*\,(1+z)^\alpha$ and
$W^*(z)=\mathcal{W}^*\,(1+z)^\beta$.  The likelihood of the data set
is then
\begin{equation}
L = \prod_i{
  \frac{(1+z_i)^{\alpha-\beta}\,e^{-\frac{W^i_0}{\mathcal{W}^*}(1+z_i)^{-\beta}}}
  {\int\int{(1+z)^{\alpha-\beta}\,e^{-\frac{W_0}{\mathcal{W}^*}(1+z)^{-\beta}}}\,
  g(W_0^{\lambda2796},z)\,dz\,dW_0} }.
\end{equation}
The parameters that maximize the likelihood are
$\alpha=0.226\pm0.170$, $\beta=0.634\pm0.097$,  and
$\mathcal{W}^*=0.443\pm0.032$\,\AA, which were determined with the aid
of the MINUIT\footnote{\copyright\ CERN, Geneva 1994-1998}
minimization software.   The maximum likelihood values are highly
correlated, and the errors include the effects of the correlations.
The resulting normalization is $\mathcal{N}^*=1.001\pm0.132$.

The full covariance matrix is
\begin{equation}
\left( \begin{array}{ccc}  \sigma^2_{\alpha\alpha} &
\sigma^2_{\alpha\beta} & \sigma^2_{\alpha \mathcal{W}^*} \\
\sigma^2_{\beta\alpha} & \sigma^2_{\beta\beta} & \sigma^2_{\beta
\mathcal{W}^*} \\ \sigma^2_{\mathcal{W}^*\alpha} &
\sigma^2_{\mathcal{W}^*\beta} & \sigma^2_{\mathcal{W}^* \mathcal{W}^*}
\\
\end{array}
\right) = \left( \begin{array}{ccc}
 0.0291 & -0.00990 &  0.00322 \\ -0.00990 &  0.00931 & -0.00292\\
 0.00322  & -0.00292 &  0.00104 \\
\end{array}
\right),
\end{equation}
which should be used when calculating the uncertainty in $\partial
^2N/\partial z\,\partial W_0$.  $\mathcal{N}^*$ was determined from
the fit so that $\int\int\,\frac{\partial ^2N}{\partial z\,\partial
W_0}\,g(W_0^{\lambda2796},z)\,dz\,dW_0$ equaled the total number of
lines in the survey.  The full covariance array was used to calculate
$\sigma_{\mathcal{N}^*}$.  No contribution from the uncertainty in the
number of lines was used, as it was estimated with a jackknife method
to be small ($\lesssim3\%$).   Since the uncertainty in
$\mathcal{N}^*$ is derived from the uncertainty in the other
parameters, it should not be considered when using equation \ref{eqn}.
For example, for $\partial N/\partial
z=\mathcal{N}^*\,(1+z)^\alpha\,e^{-\frac{W_0}{\mathcal{W}^*}(1+z)^{-\beta}}$
the uncertainty is given by
\begin{eqnarray}
\sigma^2_{\partial N/\partial z} =
\sigma^2_{\alpha\alpha}\,\left(\frac{\partial \frac{\partial
N}{\partial z}}{\partial\alpha}\right)^2 +
\sigma^2_{\beta\beta}\,\left(\frac{\partial \frac{\partial N}{\partial
z}}{\partial\beta}\right)^2 +
\sigma^2_{\mathcal{W}^*\mathcal{W}^*}\,\left(\frac{\partial
\frac{\partial N}{\partial z}}{\partial \mathcal{W}^*}\right)^2 + \\
\nonumber 2\sigma^2_{\alpha\beta}\,\left(\frac{\partial \frac{\partial
N}{\partial z}} {\partial\alpha}\right)\left(\frac{\partial
\frac{\partial N}{\partial z}}{\partial\beta}\right) +
2\sigma^2_{\alpha \mathcal{W}^*}\,\left(\frac{\partial \frac{\partial
N}{\partial z}} {\partial\alpha}\right)\left(\frac{\partial
\frac{\partial N}{\partial z}}{\partial \mathcal{W}^*}\right) +
2\sigma^2_{\beta \mathcal{W}^*}\,\left(\frac{\partial \frac{\partial
N}{\partial z}} {\partial\beta}\right)\left(\frac{\partial
\frac{\partial N}{\partial z}}{\partial \mathcal{W}^*}\right),
\end{eqnarray}
where $\frac{\partial \frac{\partial N}{\partial z}}{\partial\alpha} =
\frac{\partial N}{\partial z}\,\mathrm{ln}(1+z)$, $\frac{\partial
\frac{\partial N}{\partial z}}{\partial\beta} = \frac{\partial
N}{\partial
z}\,\frac{W}{\mathcal{W}^*}\,(1+z)^{-\beta}\,\mathrm{ln}(1+z)$, and
$\frac{\partial \frac{\partial N}{\partial z}}{\partial \mathcal{W}^*}
= \frac{\partial N}{\partial
z}\,\frac{W_0}{\mathcal{W}^{*\,2}}\,(1+z)^{-\beta}$.  All
uncertainties are statistical errors only.  Possible systematics are
discussed in \S\ref{sec_se}.

\clearpage
\begin{center}
\begin{deluxetable}{ccccccc}
\tablewidth{0pc} \tablecaption{$\partial N/\partial z$ Comparison} \tablecolumns{5}
\tablehead{ \colhead{$W_0^{\lambda2796}$} &\colhead{}&
\multicolumn{2}{c}{This Work} &\colhead{}& \multicolumn{2}{c}{SS92} \\
[1ex] \cline{3-4}\cline{6-7}\colhead{Range (\AA)}&\colhead{}&
\colhead{$\left<z_{abs}\right>$} & \colhead{$\partial N/\partial z$} &
\colhead{}&\colhead{$\left<z_{abs}\right>$} & \colhead{$\partial N/\partial z$}  }
\startdata $\ge$ 0.3 && 1.11 & 0.783 $\pm$ 0.033 && 1.12 & 0.97 $\pm$
0.10 \\
$\ge$ 0.6 && 1.12 & 0.489 $\pm$ 0.015 && 1.17 & 0.52 $\pm$ 0.07 \\
$\ge$1.0  && 1.14 & 0.278 $\pm$ 0.010 && 1.31 & 0.27 $\pm$ 0.05 \\
\enddata
\end{deluxetable}
\end{center}
\clearpage
\begin{center}
\begin{deluxetable}{ccc}
\tablewidth{0pc} \tablecaption{Summary of Data
Parameterizations\tablenotemark{a}} \tablehead{ & \colhead{$N^*$} &
\colhead{$W^*$ (\AA)}  } \startdata \multicolumn{3}{c}{Data}\\ \\  Full
Sample:         & $1.187\pm 0.052$ & $0.702 \pm 0.017$ \\ \\  $0.366
\le z \le 0.871$:  & $1.216 \pm 0.124$ & $0.585 \pm 0.024$ \\   $0.871
\le z \le 1.311$:  & $1.171 \pm 0.083$ & $0.741 \pm 0.032$ \\  $1.311
\le z \le 2.269$:  & $1.267 \pm 0.092$ & $0.804 \pm 0.034$ \\ \\
\hline\\ \multicolumn{3}{c}{Monte Carlo Simulations} \\ \\
``weak''-phase:    & $1.71 \pm 0.02$   & $0.072 \pm 0.001$ \\
``strong''-phase:  & $0.932 \pm 0.011$ & $0.771 \pm 0.014$ \\ \\
\hline\\ \multicolumn{3}{c}{Redshift Dependence} \\ \\  &
$1.001\pm0.132\,(1+z)^{0.226\pm0.170}$ &
$0.443\pm0.032\,(1+z)^{0.634\pm0.097}$ \\ \\  \enddata
\tablenotetext{a}{Maximum likelihood fits of the form:
$\partial N/\partial W_0^{\lambda2796}=\frac{N^*}{W^*}e^{-\frac{W_0}{W^*}}$.}
\end{deluxetable}
\end{center}
\clearpage
\clearpage
\clearpage
\clearpage
\clearpage
\clearpage
\clearpage
\clearpage
\clearpage
\clearpage
\clearpage
\clearpage
\clearpage
\clearpage
\clearpage
\clearpage
\clearpage
\clearpage
\clearpage
\clearpage
\clearpage


\newpage
\begin{thebibliography}{}

\bibitem[] {} Bond, N. A., Churchill, C. W., Charlton, J. C., \&  Vogt, S. S. 2001, \apj, 562, 641

\bibitem[] {} Caulet, A. 1989, \apj, 340, 90

\bibitem[] {} Charlton, J. C. \& Churchill, C. W. 1998, \apj, 499, 181

\bibitem[] {} Chen, H.-W., Lanzetta, K. M., \& Webb, J. K. 2001,
\apj, 556, 158

\bibitem[] {} Chen, H.-W., Lanzetta, K. M., Webb, J. K., \& Barcons,
X. 2001, \apj, 559, 654

\bibitem[] {} Churchill, C. W., Rigby, J. R., Charlton, J. C., \&
Vogt, S. S. 1999, \apjs, 120, 51 (CRCV99)

\bibitem[] {} Churchill, C. W., Mellon, R. R., Charlton, J. C.,
Jannuzi, B. T., Kirhakos, S,, Steidel, C. C., Schneider, D. P. 2000,
\apj, 543, 577 (C2000)

\bibitem[] {} Churchill, C. W. 2001, \apj, 560, 92

\bibitem[] {} Churchill, C. W. \& Vogt, S. S. 2001, \aj, 122, 679

\bibitem[] {} Churchill, C. W., Vogt, S. S., \& Charlton, J. C. 2003,
\aj, 125, 98

\bibitem[] {} Dickinson, M. \& Steidel, C. C. 1996, IAU Symp. 171: New
Light on Galaxy Evolution

\bibitem[] {} Driver, S. P., Fernandez-Soto, A., Couch, W. J.,
Odewahn, S. C., Windhorst, R. A., Phillips, S., Lanzetta, K., Yahil,
A. 1998, \apjl, 496, L93

\bibitem[] {} Drory, N., Bender, R., Feulner, G., Hopp, U., Maraston,
C., Snigula, J., Hill, G. J. 2003, \apj, 595, 698

\bibitem[] {} Ellison, S. L., Mall\'{e}n-Ornelas, G. \& Sawicki,
M. 2003, \apj, 589, 709

\bibitem[] {} Ellison, S. L., Churchill, C. W., Rix, S. A. \& Pettini,
M. 2004, astro-ph/0407237

\bibitem[] {} Ferguson, H. C., Dickinson, M. \& Williams, R. 2000,
\araa, 38, 667

\bibitem[] {} Guillemin, P. \& Bergeron, J. 1997, \aap, 328, 499

\bibitem[] {} Hopkins, A. M. 2004, astro-ph/0407170

\bibitem[] {} Lanzetta, K. M., Turnshek, D. A. \& Wolfe, A. M. 1987,
\apj, 322, 739

\bibitem[] {} Le Brun, V., Bergeron, J., Boiss\'{e}, P., Deharveng, J. 1997, \aap, 321, 733

\bibitem[] {} Mao, S., Mo, H. J., \& White, S. 1998, \mnras, 297, 71

\bibitem[] {} M\'{e}nard, B., Nestor, D. B. \& Turnshek, D. A. 2004,
in preparation

\bibitem[] {} Mo, H. J., \& Miralda-Escud\'{e}, J. 1996, \apj, 469, 589

\bibitem[] {} Mo, H. J., Mao, S., \& White, S. 1998, \mnras, 295, 319

\bibitem[] {} Murdoch, H. S., Hunstead, R. W., Pettini, M., \& Blades, J. C. 1986, \apj, 309, 19

\bibitem[] {} Nestor, D. B., Rao, S. M., Turnshek, D. A. \& Vanden
Berk, D. 2003, \apjl, 595, L5

\bibitem[] {} Penton, S. V., Stocke, J. T., \& Shull, J. M. 2002,
\apj, 565, 720

\bibitem[] {} Petitjean, P., \& Bergeron, J. 1990, \aap, 231, 309

\bibitem[] {} Pettini, M., Ellison, S. L., Steidel, C. C. \& Bowen,
D. V, 1999, \apj, 510, 576

\bibitem[] {} Prochaska, J. X., Gawiser, E., Wolfe, A. M., Castro,
S. \& Djorgovski, S. G. 2003, \apjl, 595, L9

\bibitem[] {} Rao, S. M. \& Turnshek, D. A. 2000, \apjs, 130, 1

\bibitem[] {} Rao, S. M., Nestor, D. B., Turnshek, D. A., Lane, W. M.,
Monier, E. M.; Bergeron, J. 2003, \apj, 595, 94

\bibitem[] {} Rao, S. M., Turnshek, D. A. \& Nestor, D. B. 2004, in
preparation

\bibitem[] {} Reed, D., Gardner, J., Quinn, T., Stadel, J., Fardal, M., Lake, G., \& Governato, F 2003, \mnras, 346, 565

\bibitem[] {} Rigby, J. R., Charlton, J. C., Churchill, C. W. 2002,
\apj, 565, 743

\bibitem[] {} Sargent, W., Young, P. J., Boksenberg, A., Tytler,
D. 1980, \apjs, 42, 41

\bibitem[] {} Sargent, W., Steidel, C. C. \& Boksenberg, A. 1988,
\apj, 334, 22

\bibitem[] {} Schneider, D. P., et al. 1993, \apjs, 87, 45

\bibitem[] {} Sheth, R. K. \& Tormen, G. 1999, \mnras, 308, 119

\bibitem[] {} Spergel, D. N. et al. 2003, \apjs, 148, 175

\bibitem[] {} Steidel, C. C. \& Sargent, W. 1992, \apjs, 80, 1

\bibitem[] {} Steidel, C. C., Dickinson, M. \& Persson, S. E. 1994,
\apjl, 437, L75

\bibitem[] {} Steidel, C. C., Kollmeier, J. A., Shapley, A. E.,
Churchill, C. W., Dickinson, M. \& Pettini, M. 2002, \apj, 570, 526

\bibitem[] {} Stoughton, C., et al. 2002, \aj, 123, 485

\bibitem[] {} Tytler, D., Boksenberg, A., Sargent, W, Young, P. \&
Kunth, D. 1987, \apjs, 64 667

\bibitem[] {} Weil, M. L., Eke, V. R., Efstathiou, G. 1998, \mnras,
300, 773

\bibitem[] {} Weymann, R. J., et al. 1998, \apj, 506, 1

\bibitem[] {} York et al. 2000, \aj, 120, 1579

\end{thebibliography}
\end{document}